\documentclass[twocolumn,pra,showpacs]{revtex4-2}
\usepackage{epsfig}
\usepackage{dcolumn}
\usepackage{natbib}
\usepackage[utf8]{inputenc}
\usepackage[english]{babel}
\usepackage{xcolor}
\usepackage{subcaption}
\usepackage{graphicx}
\usepackage{color}
\usepackage{bm}
\usepackage{tabularx}
\usepackage{amssymb}
\usepackage[intlimits]{amsmath}
\usepackage{booktabs}
\usepackage{amsmath}
\usepackage[para,online,flushleft]{threeparttable}

\definecolor{darkcandyapplered}{rgb}{0.64, 0.0, 0.0}
\usepackage[colorlinks=true,allcolors=darkcandyapplered]{hyperref}
\usepackage{physics}
\usepackage{multirow}
\usepackage{float}
\begin{document}
\title{Constraints on the Variation of Physical Constants, Equivalence Principle Violation, and a Fifth Force from Atomic Experiments}
%Yukawa-type interactions mediated by scalar particles and dark matter fields 
%\title{Variation of Fundamental Constants due to an Interaction with Ultralight Dark Matter and Constraints on the Yukawa-type Interactions mediated by Scalar Particles}
% \author{V.V. Flambaum, A. J. Mansour}
\author{V. A. Dzuba}
\email{v.dzuba@unsw.edu.au}
\author{V. V. Flambaum} 
\email{v.flambaum@unsw.edu.au}
\author{A. J. Mansour}
\email{andrew.mansour@student.unsw.edu.au}
\affiliation{School of Physics, University of New South Wales,
Sydney 2052, Australia}
\date{\today}

\begin{abstract}

The aim of this paper is to derive limits on various forms of ``new physics'' using atomic experimental data. Interactions with 
%scalar or pseudoscalar (axion)
dark energy and dark matter fields can lead to space-time variations of fundamental constants, which can be detected through atomic spectroscopy. In this study, we examine the effects of a varying nuclear mass $m_{N}$ and nuclear radius $r_{N}$ on two transition ratios: the comparison of the two-photon transition in atomic hydrogen with the hyperfine transition in $^{133}$Cs based clocks, and the ratio of optical clock frequencies in in Al$^{+}$ and Hg$^{+}$. The sensitivity of these frequency ratios to changes in $m_{N}$ and $r_{N}$ enables us to derive new limits on the variations of the proton mass, quark mass, and the QCD parameter $\theta$. Additionally, we consider the scalar field generated by the Yukawa-type interaction of feebly interacting hypothetical scalar particles with Standard Model particles in the presence of massive bodies such as the Sun and Moon. Using the data from the Al$^{+}$/Hg$^{+}$, Yb$^{+}$/Cs and Yb$^{+}$(E2)/Yb$^{+}$(E3) transition frequency ratios, we place constraints on the interaction of the scalar field with photons, nucleons, and electrons for a range of scalar particle masses. We also investigate limits on the Einstein Equivalence Principle (EEP) violating term ($c_{00}$) in the Standard Model Extension (SME) Lagrangian and the dependence of fundamental constants on gravity.  

\end{abstract}

\maketitle

\section{Introduction}

The existence of life is consistent with a narrow possible range of values for fundamental physical constants, see e.g. the review \cite{Uzan}.
This precise fitting of constants for life may be explained by the variation of these constants in space; life emerged in regions of the universe where the fundamental constants have suitable values. Potential evidence for such spatial variation in the fine structure constant $\alpha$ has been observed in quasar absorption spectra ~\cite{WebbPRL2011,MNRAS2011}. Based on this astrophysical data, it is suggested that we are moving in the direction of increasing $\alpha$, which could lead to a slow drift of fundamental constants observable in laboratory experiments ~\cite{Berengut,Berengut1}.

There are a number of models which predict the space-time variation of fundamental physical constants, which may be related to dynamical  dark energy models, dark matter models, the variation of the unification scale and string theory models - see e.g. the review \cite{Uzan}. For example, the space-time variation of fundamental constants may be due to an interaction with a slowly evolving scalar dark energy field or an oscillating dark matter field.
%One of the most interesting unsolved problems in modern physics is uncovering the nature of dark matter. Amongst other things, it is hypothesised that dark matter is made up of light bosonic particles, which are not accounted for in the standard model of elementary particles.
The dark matter candidate particles in this class are the pseudoscalar axion (and axion like particles) and the dilaton-like scalar particle~\cite{Preskill,Abbott,Dine}. If the mass of the cold dark matter is very light ($m_{\text{DM}} \ll 1 \ \text{eV}$), it may be considered to be a classical field oscillating harmonically at every particular point in space. For axions, we may write this as

\begin{align}
a = a_{0} \cos (\omega t + \varphi), \ \omega \approx m_{a} \,,
\end{align}
where $\varphi$ is a (position-dependent) phase and $m_{a}$ is the mass of the axion. Assuming that axions saturate the entire dark matter density, the amplitude $a_{0}$ may be expressed in terms of the local dark matter density $\rho_{\text{DM}} \approx 0.4 \ \text{GeV/cm}^{3}$, see e.g. Ref.~\cite{DMdensity},

\begin{align}
a_{0} = \frac{\sqrt{2 \rho_{\text{DM}}}}{m_{a}} \,.
\end{align}
Similar expressions are used to describe the case of a scalar field dark matter $\phi$.

The effects of the interaction between the scalar field and fermions may be presented as the apparent variation of fermion masses. This immediately follows from a comparison of the interaction of a fermion with the scalar field $ -g_f M_f  \phi^n \bar \psi \psi $ and the fermion mass term in the Lagrangian $- M_f \bar \psi \psi$. Adding these terms gives $M'_f=M_f(1+ g_f  \phi^n)$, with $n=1,2$. Similarly, the interaction of the scalar field with the electromagnetic field may be interpreted as a variable fine structure constant $\alpha'=\alpha (1 + g_{\gamma} \phi^n)$. These variations may manifest themselves as a slow drift or oscillations of frequencies of atomic clocks~\cite{Arvanitaki,Stadnik, Stadnik2015}. If the interaction is quadratic in $\phi$, the scalar field becomes interchangeable with the pseudoscalar (axion) field as $\phi^2$ always has positive parity~\cite{Stadnik}. The corresponding theory has been developed in Ref.~\cite{KimPerez2024}, in which limits on the axion interaction from atomic spectroscopy experiments were obtained (see also Ref.~\cite{kim2023probing,Samsonov}). 
%Note that the variation of quark and electron masses and the variation of the fine structure constant $\alpha$ are determined by different interaction constants and may be treated as independent effects.

In this paper we consider the slow drift of the physical constants, which  may be due to an interaction with an evolving non-oscillating dark energy field or the variation of the density of the dark matter field $\rho_{\text{DM}}$ in the case where $n=2$, after averaging over fast oscillations of $\cos ^2(\omega t + \varphi)=(1+ \cos2(\omega t + \varphi))/2$. Such slow variation also appears in other models~\cite{Uzan}. In then present paper we do not assume any specific model predicting the variation of the fundamental constants.  
%and fluctuations of the field amplitude $a_{0}^2$. 

 The dependence of atomic transition frequencies on $\alpha$ and the quark masses has been calculated in Refs.~\cite{PRLWebb,PRAWebb,CanJPh,Tedesco2006,Borschevsky,csquarks}. These calculations and results measuring the time dependence of atomic transition frequencies  have previously been used to place improved limits on the interaction strength of the low mass scalar field dark matter $\phi$ with photons, electrons and quarks by up to 15 orders in magnitude~\cite{Stadnik,Stadnik2}. The experimental results have been obtained by measuring the oscillating frequency ratios of electron transitions in a range of systems, including Dy/Cs \cite{DyCs}, Rb/Cs \cite{Hees2016}, Yb/Cs \cite{YbCs}, Sr/H/Si cavity \cite{HSi}, Cs/cavity \cite{Tretiak}, Cs/H~\cite{Fischer2004}, Al$^{+}$/Hg$^{+}$~\cite{AlHgdrift} and Yb$^+$/Yb$^+$/Sr \cite{Banerjee2023,Filzinger2023}.

%Cosmological and astrophysical evidence has also motivated the search for non-gravitational physics in the dark sector. An example of such physics is the existence of
Scalar particles can mediate Yukawa-type interactions between Standard Model particles.
%and produce a local variation of fundamental constants in the presence of massive bodies~\cite{Ellis1989}.
A range of experimental methods have been used to constrain these interactions, some of which include Equivalence Principle tests via Torsion Pendulum experiments~\cite{Torsion1,Torsion2}, Lunar Laser Ranging~\cite{LLRTests} and Atom Interferometry~\cite{Zhou2015}. Experimental limits on Equivalence Principle violation may be expressed in terms of the Eotvos parameter $\eta$

\begin{align}
    \eta \equiv 2 \frac{\lvert  \vec{a}_{A} - \vec{a}_{B} \rvert}{ \lvert \vec{a}_{A} + \vec{a}_{B} \rvert } \,,
\end{align}
where $\vec{a}_{A}$ and $\vec{a}_{B}$ are the accelerations of two test bodies $A$ and $B$ (which are composed of different materials) towards a third body $C$. Most recently, direct fifth force measurements have placed constraints on $\eta$. Specifically, the MISCROSCOPE mission monitored the difference in acceleration of two freely falling test masses (composed of Pt and Ti) as they orbited the Earth, constraining the Eotvos paramater to be $\eta = (-1 \pm 27) \times 10^{-15}$ at a 2-$\sigma$ confidence level~\cite{MICROSCOPE1,MICROSCOPE2}. Constraints on this parameter may be repurposed into individual limits on the couplings of the scalar particle to standard model particles.

Another method of constraining these interactions is via atomic spectroscopy  measurements. The  scalar field is produced by massive bodies and causes the variation of fundamental constants, leading to a variation in the ratio of transition frequencies. Such calculations and measurements have been performed in Refs.~\cite{Leefer2016,Brzeminski2022}.

This paper is divided into three parts. In the first part, we seek to obtain limits on the variation of physical constants. The variation of nuclear parameters affects electronic transitions, and the results of the measurements of the variation in the ratio of transition frequencies in Cs/H~\cite{Fischer2004} and the ratio of optical clocks transitions in Al$^{+}$/Hg$^{+}$~\cite{AlHgdrift} allow us to place constraints on the variation of the proton mass $m_{p}$, the variation of the average quark mass $m_{q} = (m_{u} + m_{d})/2 $ and the variation of the nuclear charge radius $r_{N}$ and nuclear mass $m_{N}$, all of which may be used to place limits on the variation of the QCD parameter $\theta$. The idea that the dependence of the electronic atomic transition frequencies on the nuclear radius (and subsequently on the hadronic parameters above) may be used in the search for dark matter fields in optical transitions was first proposed in Ref.~\cite{Banerjee2023}, while the effect of the variation of the nuclear radius (and hadronic parameters) in hyperfine transitions was earlier calculated in Ref.~\cite{Dinh2009}. In this paper, we employ a similar method to Ref.~\cite{FlambaumMansour2023}, in which the sensitivity of the optical clock transitions in Yb$^{+}$ to the variation in the nuclear radius was used to place constraints on the variation of the hadron and quark masses, and the QCD parameter $\theta$. 

In the second part we consider the beyond-standard-model effects of gravity, such as the Einstein Equivalence Principle (EEP) violating term ($c_{00}$) in the Standard Model Extension (SME) Lagrangian~\cite{Kostelecky} and the dependence of fundamental constants on the gravitational potential.

% the two photon transition in H to the ground state hyperfine transition in Cs
% , a method first proposed in Ref.~\cite{Banerjee2023}

In the third part we consider the effects produced by the interaction between hypothetical scalar particles and standard model particles in the presence of massive bodies creating a Yukawa-type potential mediated by scalar particles. Such effects have previously been considered in Ref.~\cite{Leefer2016}. The  scalar field,  produced by the Sun and the Moon, affects the fine structure constant $\alpha$ and the fermion mass $m_{f}$. We perform relevant calculations and use the measurement of the variation in the ratio of clock transition frequencies Al$^{+}$/Hg$^{+}$ in Ref.~\cite{AlHgdrift}, Yb$^{+}$/Cs in Ref.~\cite{Lange2021} and Yb$^{+}$(E2)/Yb$^{+}$(E3) in Ref.~\cite{Filzinger2023} to determine limits on the interactions between this scalar field and standard model particles for a range of scalar particle masses. These constraints are compared to those obtained from the results of the MICROSCOPE mission~\cite{MICROSCOPE1,MICROSCOPE2}.

% We are also consider other interpretations of such experiments.

% Light scalar fields appear very naturally in modern cosmological models, and massive bodies such as stars and galaxies may act as sources of these fields due to their charge producing a Coulomb-like scalar field.

Note that when discussing the variation of dimensionful parameters, we must be mindful of the units they are measured in, as these units may also vary. In other words, we should consider the variation of dimensionless parameters which have no dependence on any measurement units.  Nuclear properties depend on the quark mass and $\Lambda_{\text{QCD}}$. In this work, we keep $\Lambda_{\text{QCD}}$ constant, meaning our calculations are related to the measurement of the variation of a dimensionless parameter $X_q \equiv m_q/\Lambda_{\text{QCD}}$. As a result, we measure the quark mass in units of $\Lambda_{\text{QCD}}$ - see Refs.~\cite{FlambaumShuryak,Wiringa2009,Tedesco2006}. A similar choice of units is assumed for the variation of hadron masses.

In this paper, we assume natural units $\hbar=c=1$ if $\hbar$ and $c$ are not explicitly presented.

\section{Atomic transition frequency shift due to variation of nuclear mass and radius}
The total electronic energy $E_{\text{tot}}$ of an atomic state contains the energies associated with the finite mass of the nucleus $m_{N}$ (mass shift, MS) and the non-zero nuclear charge radius $r_{N}$ (field shift, FS). Both effects contribute to the isotope shifts of atomic transition frequencies. These are parametrised as~\cite{Krane}

\begin{align} \label{Energies}
    E_{\text{MS}} \approx K_{\text{MS}} \frac{1}{m_{A}} \propto \frac{1}{A} \ \text{and} \ E_{\text{FS}} \approx K_{\text{FS}} r_{N}^{2} \propto A^{2/3} \,,
\end{align}
where $K_{\text{MS}}$ and $K_{\text{FS}}$ are the mass and field shift coefficients respectively and $m_{A}$ is the mass of an atom with atomic mass number $A$, which is largely determined by the nuclear mass $m_{N}$. The variation of the total electronic energy associated with the nuclear degrees of freedom can be written as~\cite{king2013isotope}

\begin{align}
    \frac{\delta E_{\text{tot}}}{E_{\text{tot}}} = - \frac{E_{\text{MS}}}{E_{\text{tot}}} \frac{\delta m_{N}}{m_{N}} + \frac{E_{\text{FS}}}{E_{\text{tot}}} \frac{\delta r_{N}^{2}}{r_{N}^{2}} \,.
\end{align}
The mass shift term dominates for light nuclei whilst the field shift term dominates for heavy nuclei. In general, comparing the electronic transition frequencies $\nu_{a}$ and $\nu_{b}$ of two different atomic species, we obtain

\begin{align} \label{variation}
    \frac{\delta (\nu_{a} / \nu_{b})}{(\nu_{a} / \nu_{b})} = (K_{1} - K_{2}) \frac{\delta r_{N}^{2}}{r_{N}^{2}}  + (K_{3} - K_{4}) \frac{\delta m_{N}}{m_{N}} \,,
\end{align}
where

\begin{align}
    K_{1} & = \frac{K_{\text{FS}}^{\nu_{a}} r_{N,a}^{2} }{\nu_{a}} \,, \\
    K_{2} & = \frac{K_{\text{FS}}^{\nu_{b}} r_{N,b}^{2} }{\nu_{b}} \,, \\
    K_{3} & =   \frac{K_{\text{MS}}^{\nu_{a}}}{\nu_{a} m_{N,a}} \,, \\
    K_{4} & = \frac{K_{\text{MS}}^{\nu_{b}}}{\nu_{b} m_{N,b}} \,.
\end{align}

\section{Limits on the linear drift of the QCD parameter $\theta$ and particle masses}
In this section, we use the above theory along with experimental observations of the drift in atomic transition frequencies to place limits on the variation of quark and hadronic parameters. The approach in this section is similar to that presented in Ref.~\cite{FlambaumMansour2023}. 
% Standard model spinor fields $\psi$, photon $F_{\mu\nu}$ and gluon $G^l_{\mu\nu}$ fields can have the following interaction vertices with a pseudoscalar field $a$:
% \begin{equation}
% \label{V1}
% V= \frac{C_f}{f_a}  \partial_{\mu}  a  \bar \psi  \gamma_5 \gamma^{\mu}\psi + 
%  C_{\gamma} \frac{a}{f_a} \tilde{F}^{\mu\nu} F_{\mu\nu} +  C_g \frac{a}{f_a} \tilde{G}^{l\,\mu\nu} G^l_{\mu\nu}\,.  
% \end{equation}
% Here $C_f$, $C_\gamma$ and $C_g$ are dimensionless constants which are of order $O(1)$ for the QCD axion model, but are arbitrary for the general pseudoscalar (axion-like) particle. In particular, upon the substitution $C_g = g^2/(32\pi^2)$, or
% \begin{equation}
% \label{theta}
%     \theta = \frac{32\pi^2 C_g a}{g^2 f_a}\,
% \end{equation}
% the last term in Eq.~(\ref{V1}) reduces 
% to %the QCD $\theta$-term (\ref{thetaQCD}) 
% the standard QCD $\theta$-term
% \begin{equation}
% \label{thetaQCD}
%     \frac{g^2\theta}{32\pi^2}\tilde G^{l\,\mu\nu}G^l_{\mu\nu}\,,
% \end{equation}
% where $\theta=a/f_a$, $f_a$ is the axion decay constant, $g$ is the strong interaction coupling constant, $G^l_{\mu\nu}$ is the gluon field strength and $\tilde G^{l\,\mu\nu}$ is its dual.
% Thus, the classical axion dark matter field $a=a_0 \cos (m_a t + \varphi)$ may be interpreted as a dynamical QCD parameter $\theta=a/f_a$.

Let us first consider the variation of fundamental constants due to the variation of the ratio of transition frequencies in Cs/H. Ref.~\cite{Fischer2004} compared results of the variation in the $\ket{1^{2}S_{1/2}, F=1 , m_{F} = \pm 1 } \rightarrow \ket{2^{2}S_{1/2}, F^{\prime} = 1, m_{F}^{\prime} = \pm 1}$ two-photon transition in atomic hydrogen to results from clocks based on $^{133}$Cs in order to deduce limits on the fractional time variation of the fine structure constant $\alpha$. The comparison of the transition in H against the ground state hyperfine transition in $^{133}$Cs gives the following fractional time variation~\cite{Fischer2004}

\begin{align} \label{CsHLimit}
    \frac{1}{(\nu_{\text{Cs}}/\nu_{\text{H}}) } \frac{d (\nu_{\text{Cs}}/\nu_{\text{H}})}{dt} =  3.2 \ (6.3) \times 10^{-15} \ \text{yr}^{-1} \,.
\end{align}
The variation of this ratio may be related to the variation of fundamental constants - see Ref.~\cite{Tedesco2006}

\begin{align}
    \frac{\delta (\nu_{\text{Cs}}/\nu_{\text{H}})}{(\nu_{\text{Cs}}/\nu_{\text{H}}) } = 2.83 \frac{\delta \alpha}{\alpha} + 0.009 \frac{\delta m_{q}}{m_{q}} + \frac{\delta (m_{e} /m_{p})}{(m_{e}/m_{p})} \,.
\end{align}
The relative variation of the electron to proton mass ratio can be described as~\cite{Flambaum2004}

\begin{align}
    \frac{\delta (m_{e} /m_{p})}{(m_{e}/m_{p})} = -0.037 \frac{\delta m_{q}}{m_{q}} - 0.011 \frac{\delta m_{s}}{m_{s}} + \frac{\delta m_{e}}{m_{e}} \,,
\end{align}
where $m_{s}$ is the mass of the strange quark. For brevity, we may assume that $\delta m_{s}/m_{s} = \delta m_{q}/m_{q}$. Combining these expressions gives 

\begin{align}
    \frac{\delta (\nu_{\text{Cs}}/\nu_{\text{H}})}{(\nu_{\text{Cs}}/\nu_{\text{H}}) } = 2.83 \frac{\delta \alpha}{\alpha} -0.039 \frac{\delta m_{q}}{m_{q}} + \frac{\delta m_{e}}{m_{e}} \,.
\end{align}
Therefore, using this expression along with the limit presented in Eq. (\ref{CsHLimit}), we may obtain a limit on the variation of the quark mass $m_{q}$ assuming that there is no variation of $\alpha$ and $m_e$

\begin{align}
    \frac{1}{m_{q}} \frac{d m_{q}}{dt} = -8.2 \ (16) \times 10^{-14} \ \text{yr}^{-1} \,. \label{variationquark1}
\end{align}
When considering the variation of the QCD parameter $\theta$, it is convenient to consider the problem at the hadron level, rather than the quark level. Using the calculations presented in Ref.~\cite{Tedesco2006} and the limit (\ref{CsHLimit}) we obtain 

%As such, we may use this relation to place limits on the variation of the pion mass using the following result presented in Refs.~\cite{Flambaum2006,Holl2006} 
%\begin{align}
%    \frac{\delta m_{\pi}}{m_{\pi}} & = 0.498  \frac{\delta m_{q}}{m_{q}} \,. \label{variationpionquark}
%\end{align}
%Therefore, using Eqs. (\ref{variationquark1}, \ref{variationpionquark}) we estimate the variation of the pion mass 
\begin{align}
    \frac{\delta m_{\pi}}{m_{\pi}} & = -4.1 \ (8.0) \times 10^{-14} \ \text{yr}^{-1} \,. \label{pion2Eq}
\end{align} 
Using this value, along with the following result from Ref.~\cite{KimPerez2024}

\begin{align} \label{pionproton}
    \frac{\delta m_{p}}{m_{p}} = 0.13 \frac{\delta m_{\pi}}{m_{\pi}}\,,
\end{align}
we may subsequently place limits on the drift of the proton mass due to the drift of the pion mass

\begin{align}
    \frac{1}{m_{p}} \frac{d m_{p}}{dt}= -5.3 \ (10) \times 10^{-15} \ \text{yr}^{-1} \,.
\end{align}

Finally, we may use the relation between the variation of the pion mass and the QCD parameter $\theta$, presented in Ref.~\cite{Ubaldi2010}, to place limits on the linear drift of $\theta$. The pion mass has dependence on $\theta$, and the shift of the pion mass due to a small $\theta$ relative to the pion mass for $\theta=0$ is given  by~\cite{Ubaldi2010}:

\begin{align}
    \frac{\delta m_{\pi}}{m_{\pi}} & = -0.05 \ \theta^{2} \, \label{ThetaDef}
\end{align}
Thus, substituting the value for the drift of the pion mass from Equation (\ref{pion2Eq}) yields
\begin{align}
    \frac{d \theta^{2}}{dt} & = 8.2 \ (16) \times 10^{-13} \ \text{yr}^{-1} \,.
\end{align}

% \subsection{Variation of the quark mass due to the drift of the $\text{Al}^{+}/\text{Hg}^{+}$ frequency ratio}
We may also perform a similar calculation relevant to measurements of a different ratio of transition frequencies. The drift of the ratio of the optical clock transition frequencies of aluminium and mercury has been measured in Ref.~\cite{AlHgdrift}. This drift was used to place limits on the temporal variation of the fine structure constant. Using a similar method to that above, this result may be repurposed to extract limits on the variation of the nuclear radius and hadronic parameters. The rate of change in the ratio of the transition frequencies of $^{1}S_{0} \rightarrow \ ^{3}P_{0}$ transitions in Al$^{+}$ and $^{2}S_{1/2} \rightarrow \ ^{2}D_{5/2}$ transitions in Hg$^{+}$ was found to be~\cite{AlHgdrift}

\begin{align} \label{al/hgfreq}
    \frac{1} {{\nu_{\text{Al}^{+}}}/ \nu_{\text{Hg}^{+}}} \frac{d ( \nu_{\text{Al}^{+}} / \nu_{\text{Hg}^{+}})}{dt} = -5.3 \ (7.9) \times 10^{-17} \ \text{yr}^{-1} \,.
\end{align}
Once again we note the fact that mass shift dominates the isotope shift effects in light elements, while field (volume) shift dominates in heavy elements. As such, the isotopic shift in $\text{Al}^{+}$ is dominated by the mass shift component, while the isotopic shift in $\text{Hg}^{+}$ is dominated by the field shift component. 

The mass shift is divided into two components: the normal mass shift (NMS) and the specific mass shift (SMS). The NMS results from a change in the reduced electron mass, and its parameter is easily calculated from the transition frequency using the non-relativistic virial theorem stating that in the case of the Coulomb interaction, the electron's total energy change corresponds to the change in it's kinetic energy, with a negative sign:

% the total electron energy change is equal to the electron kinetic energy change with a  negative sign:

\begin{align}
    K_{\text{NMS}}^{\nu} = -\frac{\nu}{1822.888} \ \text{amu}\,, 
\end{align}
where the factor in the denominator refers to the ratio of the atomic mass unit to the electron mass. %The SMS, which arises in multi-electron atoms and ions, is difficult to evaluate accurately. 
Using the transition frequency from the experimental observation of the $^{1}S_{0} \rightarrow \ ^{3}P_{0}$ transition in $^{27}$Al$^{+}$ of Ref.~\cite{Wineland2007}, we calculate the normal mass shift factor in Al$^{+}$ to be $K_{\text{NMS}}^{\nu_{\text{Al}^{+}}} = -615 $ GHz amu. Using the calculated ratio of the specific to normal mass shifts from Ref.~\cite{Tang2021}, 
%This result for $K_{\text{NMS}}^{\nu_{\text{Al}^{+}}} $ differs slightly to the results of the theoretical calculations presented in~\cite{Tang2021}, in which the mass shift factors are determined to be $K_{\text{NMS}}^{\nu_{\text{Al}^{+}}} = -654$ GHz amu and $K_{\text{SMS}}^{\nu_{\text{Al}^{+}}} = -972$ GHz amu. From these values, we note that the ratio of SMS and the NMS parameter for the $^{1}S_{0} \rightarrow \ ^{3}P_{0}$ transition in Al$^{+}$ is $\sim 1.5$. Using this ratio along with the calculated value of the NMS parameter $K_{\text{NMS}}^{\nu_{\text{Al}^{+}}} = -615$ GHz amu, we approximate a corrected value of the SMS parameter in Al$^{+}$ to be $K_{\text{SMS}}^{\nu_{\text{Al}^{+}}} = -914$ GHz amu. Therefore, 
we yield a total mass shift parameter of $K_{\text{MS}}^{\nu_{\text{Al}^{+}}} = -1530 $ GHz amu.

% Al$^{+}$ has the same electronic structure as Mg. The isotope shift parameters for Mg were calculated in Ref~\cite{Berengut2005}. As such, we may use the results of their calculation to find the ratio of SMS and the NMS parameter for the $^{1}S_{0} \rightarrow \ ^{3}P_{0}$ transition in Mg, which may be used to approximate the SMS parameter in Al$^{+}$.

Now we must consider the effects from the field shift in Hg$^{+}$. 
We have performed relativistic many-body calculations of the shift of atomic transition frequencies due to the variation of the nuclear radius in Hg$^{+}$. The wave functions and energies in the zeroth-order approximation have been calculated using the Dirac-Hartree-Fock method, including Breit corrections. The correlation corrections were calculated using the CIPT method (configuration interaction with perturbation theory~\cite{CIPT}), which allows us to deal with open electronic shells. This approach is similar to that  described in Ref.~\cite{NoIS}. We have found the field shift parameter for the $^{2}S_{1/2} \rightarrow \ ^{2}D_{5/2}$ transition in Hg$^{+}$ to be $85.5 $ GHz/fm$^{2}$. Similar calculations are performed to find isotope shifts in atomic transitions.
Our value for the isotope shift in the $^{2}S_{1/2} \rightarrow \ ^{2}D_{5/2}$ transition of Hg$^{+}$ is in agreement with the calculations of the isotope shift in Hg$^{+}$ presented in Ref.~\cite{Zhang2019}.
Thus, using the MS and FS parameters along with Eq. (\ref{variation}), we obtain
%we may extract independent limits on the variation of fundamental parameters

\begin{align}
\begin{split}
    \frac{\delta ( \nu_{\text{Al}^{+}} / \nu_{\text{Hg}^{+}})} {{\nu_{\text{Al}^{+}}}/ \nu_{\text{Hg}^{+}}} = - \bigg[  \frac{K_{\text{FS}}^{\nu_{\text{Hg}^{+}}} r_{N,\text{Hg}^{+}}^{2} }{\nu_{\text{Hg}^{+}}} \frac{\delta r_{0}^{2}}{r_{0}^{2}} + \\ \frac{K_{\text{MS}}^{\nu_{\text{Al}^{+}}}}{\nu_{\text{Al}^{+}} m_{N,\text{Al}^{+}}} \frac{\delta m_{N}}{m_{N}} \bigg] \,.
\end{split} \label{AlHgVariationRatio}
\end{align}
Here, we have used the fact that the nuclear radii in all nuclei may be quite accurately related to the internucleon distance $r_{0}$ by the universal formula $r_{N} = A^{1/3} r_{0}$. This implies that these quantities have equivalent fractional variations. Presenting limits on the variation of $r_{0}$ is more useful as it allows one to compare the results of measurements in different nuclei.  
Noting that $r_{N,\text{Hg}^{+}}^{2} = 5.4474$ fm in $^{199}$Hg~\cite{Angeli2013} and $m_{N, \text{Al}^{+}} \approx 27 m_p$  in $^{27}$Al, we yield the following limits
\begin{align} 
\frac{1}{r_{0}} \frac{d r_{0}}{dt} & = -1.1 \ (1.7) \times 10^{-14} \ \text{yr}^{-1} \,, \label{Limitsr0} \\
     \frac{1}{m_p} \frac{d m_{p}}{dt} & = 1.0 \ (1.6) \times 10^{-12} \ \text{yr}^{-1} \,. \label{Limitsmn}
\end{align}
%We note the fact that these effects have opposing sign, meaning they may partially cancel each other out. 
Considering these effects as separate sources of the variation allows us to derive independent limits on the variation of quark and hadronic parameters.

Firstly, using a similar method to the one presented in Ref.~\cite{FlambaumMansour2023}, we consider the effects arising from a variation in the internucleon distance $r_{0}$. Calculations of the dependence of nuclear energy levels and nuclear radii on fundamental constants were performed in Refs.~\cite{FlambaumShuryak,Wiringa2007,Wiringa2009}. Specifically, in Table VI of Ref.~\cite{Wiringa2009}, the sensitivity coefficients of nuclear radii to the variation of hadron masses for several light nuclei have been presented. These results may be extended to all nuclei due to the relation $r_N=A^{1/3} r_0$. This relation follows from the constancy of nuclear density and reasonably describes the nuclear radius for nuclei with mass number $A>2$. %This implies that fractional variations in the nuclear radius $r_N$ and the internucleon distance $r_0$  are equivalent. 
Therefore, by calculating the variation in $r_0$, a fundamental parameter, we can generalize the results for light nuclei to include all nuclei. 
The sensitivity coefficients are defined by the relation
\begin{align} \label{r0}
    \frac{\delta r_{0}}{r_{0}} = \sum_h K_{h} \frac{\delta m_{h}}{m_{h}} \,.
\end{align}
The sum over hadrons in Refs.~\cite{Wiringa2007,Wiringa2009}
includes contributions from $\pi$, nucleon, $\Delta$ and vector mesons (these hadron masses are parameters of the kinetic energy and nucleon interaction operators used in Refs.~\cite{Wiringa2007,Wiringa2009}). The sensitivity to the pion mass is given by the coefficient $K_{\pi} = 1.8$ and the sensitivity to the nucleon mass is given by $K_{n} = -4.8$. We neglect contributions from $\Delta$ and vector mesons as they are of a similar magnitude with opposing sign, meaning their resulting contribution is small and unstable. Note that the estimate in Eq. (\ref{r0}) is model dependent and may have  an error O(1).

%and the sensitivity to vector mesons is given by $K_{V} = -12.9$. 
Subsequently, the variation of hadron masses may be related to variation of the quark mass, see e.g.  Ref.~\cite{Roberts2008}:
\begin{equation}
  \frac{\delta m_h}{m_h} = K_{h,q} \frac{\delta m_q}{m_q}\,,
\end{equation}
where $m_{q} = (m_{u} + m_{d})/2$ corresponds to the average light quark mass. The sensitivity coefficient for the  pion mass is an order of magnitude bigger than that for other hadrons since the pion mass vanishes for zero quark mass ($m_{\pi} \propto m_q^{1/2}$) while other hadron masses remain finite. Indeed, according to Refs.~\cite{Flambaum2006,Holl2006} $K_{\pi,q}=0.498$ for the pion, while $K_{n,q}= 0.06$ for nucleons. The sensitivity coefficients to the quark mass have been calculated for light nuclei in Ref.~\cite{Wiringa2009}. The average value is given by 
\begin{align} \label{dr}
    \frac{\delta r_{0}}{r_{0}} = 0.3 \frac{\delta m_{q}}{m_{q}} \,.
\end{align}
We note that here there are partial cancellations of different contributions, so the sensitivity is smaller than that following from pion mass alone. Refs.~\cite{Wiringa2007,Wiringa2009} have also presented calculations of the dependence of the nuclear energies and radii on the variation of the fine structure constant $\alpha$. Applying Equation (\ref{dr}) to the limit on the variation of the internucleon distance presented in (\ref{Limitsr0}), we determine limits on the variation of the quark mass to be

\begin{align}
    \frac{1}{m_{q}} \frac{d m_{q}}{dt}= -3.7 \ (5.7) \times 10^{-14} \ \text{yr}^{-1} \,.
\end{align}
Once again, in order to place limits on the variation of the proton mass and the QCD parameter $\theta$, it is convenient to consider the problem at the hadron level, rather than the quark level. In Ref.~\cite{Wiringa2009}, the sensitivity of the nuclear radius to the masses of the pion, nucleon, vector meson and delta has been calculated. In the following estimate, we do not include contributions from the vector meson and delta as their contributions are smaller. These contributions also have opposing signs, meaning they partially cancel each other out making their contribution less reliable. The variation of the nuclear radius may be written in terms of the pion and nucleon mass as
\begin{align} \label{rpi}
 \frac{\delta r_0}{r_0}=1.8  \frac{\delta m_{\pi}}{m_{\pi}} -  4.8 \frac{\delta m_n}{m_n} = 1.2   \frac{\delta m_{\pi}}{m_{\pi}} \,,
\end{align}
where in the last equality we have applied Equation (\ref{pionproton}). Thus, once again applying Equation (\ref{Limitsr0}), we obtain a limit on the drift of the pion mass

\begin{align}
    \frac{1}{m_{\pi}} \frac{d m_{\pi}}{dt} = -9.2 \ (14) \times 10^{-15} \ \text{yr}^{-1} \,.
\end{align}
Finally, we apply Equation (\ref{ThetaDef}) in order to place constraints on the linear drift of the QCD parameter $\theta$

\begin{align}
    \frac{d \theta^{2}}{dt} = 1.8 \ (2.8) \times 10^{-13} \ \text{yr}^{-1} \,.
\end{align}
These limits are approximately 44 times weaker than the limits imposed from the sensitivity of the optical clock transitions in Yb$^{+}$ obtained in Ref.~\cite{FlambaumMansour2023}. This difference corresponds exactly to the difference in the accuracy of the experimental measurements of the variation in clock frequency ratios, showing that these systems have equivalent sensitivities to the variation of the internucleon distance. We however note that the limits from the Al$^{+}$/Hg$^{+}$ clock ratio are $\sim 2-4$ times stronger than those calculated for the Cs/H system, despite the fact that the accuracy of the Al$^{+}$/Hg$^{+}$ measurements is $\sim 60$ times better. This implies that the sensitivity of the Cs/H ratio to changes in the hadron  constants is higher than that of the Al$^{+}$/Hg$^{+}$ ratio. 

Let us now consider the effects arising from a variation of the nuclear mass $m_{N}$. The nuclear mass may be related to the proton mass $m_{p}$ by the following relation $m_{N} \approx A \ m_{p}$. As such, these parameters have equivalent fractional variations, meaning the limit from Eq. (\ref{Limitsmn}) applies. Using this result, along with the relation between the pion and the proton mass presented in Equation (\ref{pionproton}), we may place limits on the drift of the pion mass

\begin{align}
     \frac{1}{m_{\pi}} \frac{d m_{\pi}}{dt} = 8.1 \ (12) \times 10^{-12} \ \text{yr}^{-1} \,.
\end{align}
Using the calculations presented in Ref.~\cite{Tedesco2006}, we may use this result to place limits on the variation of the quark mass $m_{q}$
% We may now use the relation between the variation of the pion mass and the quark mass $m_{q}$ presented in Equation (\ref{variationpionquark}) to place limits on the variation of $m_{q}$
\begin{align}
\begin{split}
    \frac{1}{m_{q}} \frac{d m_{q}}{dt} & = 1.6 \ (2.4) \times 10^{-11} \ \text{yr}^{-1} \,.
\end{split}
\end{align}
We also use the relation between the variation of the pion mass and the QCD parameter $\theta$ presented in Equation (\ref{ThetaDef}) to place constraints on the linear drift of $\theta^{2}$
\begin{align}
\begin{split}
    \frac{d \theta^{2}}{dt} & = -1.6 \ (2.4) \times 10^{-10} \ \text{yr}^{-1} \,.
\end{split}
\end{align}
As expected, the limits obtained upon considering the variation in the ratio of frequencies (\ref{al/hgfreq}) as being due to the drift of the nuclear (and hence nucleon) mass are weaker than the limits from the variation of the internucleon distance.  

\section{Gravity related variation of fundamental constants}

In some theoretical models, atomic transition frequencies and fundamental constants may depend on the gravitational potential. In Ref.~\cite{Flambaum2007} limits on the gravity related variation of fundamental constants are derived from measurements of the drift of atomic clock frequency ratios. In a similar way, we may obtain limits on the gravity related variation of the nuclear radius and nuclear mass using the measurement of the variation in the ratio of transition frequencies in $\text{Al}^{+}/\text{Hg}^{+}$ from Refs.~\cite{Dzuba2018,AlHgdrift}. We will also calculate limits on gravity related variation of these quantities for other systems of interest, Yb$^{+}$/Cs~\cite{Lange2021} and Yb$^{+}$/Yb$^{+}$~\cite{Filzinger2023}.

%also expect fundamental constants to have a dependence on the position of the measurement point relative to the source of the potential. Therefore, 

Noting the dependence of the frequency shift on the nuclear mass and radius from Eq. (\ref{variation}), we introduce the parameters $\kappa_{r,N}$ as follows: (see Ref.~\cite{Flambaum2007})

\begin{align}
    \frac{\delta r_{N}^{2}}{r_{N}^{2}} & = \kappa_{r} \ \delta \left( \frac{GM}{rc^2} \right) \,, \label{rgravity} \\
    \frac{\delta m_{N}}{m_{N}} & = \kappa_{N} \ \delta \left( \frac{GM}{rc^2} \right)\,. \label{mgravity}
\end{align}
It is instructive to link these parameters  $\kappa_{r}$ and $\kappa_{N}$ to the Einstein Equivalence Principle (EEP) violating term in the Standard Model Extension Hamiltonian ~\cite{Kostelecky}.
This term may be  presented as a correction to the kinetic energy which in non-relativistic form is equal to (see e.g.~\cite{Haffner})
\begin{equation}
\delta H=\frac{2}{3}c_{00}\frac{U}{c^2}\frac{p^2}{2m_{e}}, 
\end{equation} 
where $c_{00}$ is a parameter of the Standard Model Extension (SME) Lagrangian  \cite{Kostelecky}, $U$ is the gravitational potential, $p$ is the electron momentum operator and $m_{e}$ is the electron mass. Limits on $c_{00}$ have been found by monitoring the drift of atomic clock frequencies for different transitions (see e.g.~\cite{Dzuba2018,Haffner}). Ref.~\cite{Dzuba2018} subsequently used the limits on this parameter to determine a limit on the gravity-related variation of the fine structure constant $\kappa_{\alpha}$
\begin{align} \label{alphagravity}
    \frac{\delta \alpha}{\alpha} = \kappa_{\alpha} \ \delta \left( \frac{GM}{c^2r} \right) \,,
\end{align}
where in the case of a relative change of two transition frequencies in different atomic species $\delta \omega_{a} / \omega_{a} = K_{\alpha a} (\delta \alpha / \alpha)$ and $  \delta \omega_{b} / \omega_{b} = K_{ \alpha b} (\delta \alpha / \alpha) $, the parameter $\kappa_{\alpha}$ is equal to 

\begin{align}
    \kappa_{\alpha} = \frac{2}{3} \frac{R_{b} - R_{a}}{K_{\alpha a} - K_{\alpha b}} c_{00} \,,
\end{align}
where %$c_{00}$ is the parameter characterising the magnitude of the EEP violation. 
$R_{a}$ and $R_{b}$ represent the relativistic factors of atoms $a$ and $b$ respectively, which describe the deviation from the expectation value of the kinetic energy of a relativistic atomic electron from the value given by the non-relativistic virial theorem~\cite{Dzuba2018}

\begin{align}
    R = - \frac{\Delta E_{i} - \Delta E_{j}}{E_{i} - E_{j}} \,,
\end{align}
where $\Delta E_{i}$ is the energy shift of the state $i$ due to the relativistic kinetic energy operator. In the non-relativistic limit, $R=1$. Thus, we employ a similar method and derive expressions for the parameters $\kappa_{r}$ and $\kappa_{N}$ describing the variation of the nuclear radius and nuclear mass:

\begin{align}
    \kappa_{r} & =  \frac{2}{3} \frac{R_{b} - R_{a}}{K_{1} - K_{2}} c_{00} \,, \label{kappar} \\
    \kappa_{N} & =  \frac{2}{3} \frac{R_{b} - R_{a}}{K_{3} - K_{4}} c_{00} \,, \label{kappam} 
\end{align}
where $K_{1,2,3,4}$ are defined in Eq. (\ref{variation}). The value of $R$ has been calculated for a number of clock transitions in Ref.~\cite{Dzuba2018}, see Table \ref{RTable}. Ref.~\cite{Dzuba2018} also calculated a limit on the value of the SME parameter $c_{00}$ due to the change in the Sun's gravitational potential for the ratio of these transition frequencies $\nu_{\text{Al}^{+}}/\nu_{\text{Hg}^{+}}$. As such, we may substitute all known quantities into Eqs. (\ref{kappar}, \ref{kappam}) and determine a limit on the gravity related variation of the nuclear radius and nuclear mass. We also use the limits on the coupling to gravity of the fine structure constant $\alpha$, presented in Refs. \cite{Lange2021} (Yb$^{+}$/Cs) and \cite{Filzinger2023} (Yb$^{+}$/Yb$^{+}$) in order to place improved limits on gravity's coupling to the nuclear radius $\kappa_{r}$. In performing these calculations, we make use of the following result for the difference in field shift parameters in the electric octupole (E3) and electric quadrupole (E2) transitions in Yb$^{+}$: $ K_{1} - K_{2} = 2.4 \times 10^{-3}$~\cite{Banerjee2023}, and the sensitivity of variations in this ratio to the fine structure constant $\alpha$, $k_{\alpha}(E3) - k_{\alpha}(E2) = -6.95$~\cite{CanJPh}. The results are presented in Table \ref{Kappatable}.

% In the following subsections, we use these parameters to determine limits on the Yukawa-type interaction of the scalar field $\phi_{s}$ from the Sun and the Moon with Standard Model particles.

% and the electron to proton mass ratio $m_{e}/m_{p}$ presented in Ref.~\cite{Lange2021} (Yb$^{+}$/Cs) to calculate the 

% where $K_{1,2,3,4}$ are defined in Eq. (\ref{variation}). The value of $R$ has been calculated for a number of clock transitions in Ref.~\cite{Dzuba2018}. Specifically, we note that $R_{\text{Al}^{+}} (^{1}S_{0} \rightarrow \ ^{3}P_{0}) = 1$ and $R_{\text{Hg}^{+}} (^{2}S_{1/2} \rightarrow \ ^{2}D_{5/2}) = 0.2$. Ref.~\cite{Dzuba2018} also calculated the value of $c_{00}$ due to the change in the Sun's gravitational potential for the ratio of these transition frequencies, $\nu_{\text{Al}^{+}}/\nu_{\text{Hg}^{+}}$, by fitting the results of the measurements performed in Ref.~\cite{AlHgdrift}. Using their calculated value of $c_{00} = -3.0 \ (5.7) \times 10^{-7}$, we may substitute all known quantities into Eqs. (\ref{kappar}, \ref{kappam}) and determine a limit on the gravity related variation of the nuclear radius and nuclear mass respectively:

% In performing these calculations, we make use of the following result for the difference in field shift parameters in the electric octupole (E3) and electric quadrupole (E2) transitions in Yb$^{+}$: K_{1} - K_{2} = 2.4 \times 10^{-3}~\cite{Banerjee2023}, and the sensitivity of variations in this ratio to the fine structure constant $\alpha$ $k_{\alpha} = -6.95$~\cite{CanJPh}. 

\begin{table}
\begin{center}
\begin{tabular}{ccccc}
\hline \hline 
Atom/Ion & Ground state  &  Clock state  & $\hbar \omega\left[\mathrm{cm}^{-1}\right]$ & $R$ \\
\hline $\mathrm{Al}^{+}$ & $3 s^2$ \  ${ }^1 \mathrm{S}_0$ & $3 s 3 p$  \ ${ }^3  \mathrm{P}_0^o$ & 37393 & 1.00 \\
$\mathrm{Hg}^{+}$ & $5 d^{10} 6 s$  ${ }^2 \mathrm{S}_{1 / 2}$ & $5 d^9 6 s^2$  ${ }^2 \mathrm{D}_{5 / 2}$ & 35515 & 0.2 \\
$\mathrm{Yb}^{+}$ & $6 s$ \  ${ }^2 \mathrm{S}_{1 / 2}$ & $5 d$ \ ${ }^2  \mathrm{D}_{3 / 2}$ & 22961 & 1.48 \\
$\mathrm{Yb}^{+}$ & $6 s$ \  ${ }^2 \mathrm{S}_{1 / 2}$ & $4 f$ \  ${ }^2 \mathrm{F}_{7 / 2}$ & 21419 & -1.9 \\

\hline \hline
\end{tabular}
\caption{Relativistic factors ($R$) for optical clock transitions in atoms and ions~\cite{Dzuba2018}.}
\label{RTable}
\end{center}
\end{table}

%\begin{table*}[tbh]
%\begin{center}
%\begin{tabularx}{\textwidth}{  >{\centering\arraybackslash}X >{\centering\arraybackslash}X >{\centering\arraybackslash}X >{\centering\arraybackslash}X >{\centering\arraybackslash}X}
%\hline \hline 
%System &  $c_{00}$ & Ref. & $\kappa_{r}$  &  $\kappa_{N}$  \\
%\hline $\mathrm{Al}^{+}/\mathrm{Hg}^{+}$ & $-3.0 \ (5.7) \times 10^{-7} $ & \cite{Dzuba2018} & $-6.7 \ (13) \times 10^{-5}$ & $-3.1 \ (6.0) \times 10^{-3}$ \\
%$\mathrm{Yb}^{+}/\mathrm{Cs}$  & $4.2 \ (3.3) \times 10^{-8}$ & [This work] & $4.0 \ (3.1) \times 10^{-5}$ & $7 \ (45) \times 10^{-8}$\\
%$\mathrm{Yb}^{+}/\mathrm{Yb}^{+}$ &  $-7.4 \ (9.3) \times 10^{-9}$ & [This work] & $7.0 \ (8.7) \times 10^{-6}$ & -- \\
%\hline \hline
%\end{tabularx}
%\caption{Limits on SME parameter $c_{00}$ and the gravity related variation of the nuclear charge radius $\kappa_{r}$ and the nuclear mass $\kappa_{N}$. }
%\label{Kappatable}
%\end{center}
%\end{table*}
% \begin{tabularx}{\textwidth}{  >{\centering\arraybackslash}X >{\centering\arraybackslash}X >{\centering\arraybackslash}X >{\centering\arraybackslash}X >{\centering\arraybackslash}X >{\centering\arraybackslash}X >{\centering\arraybackslash}X }

\begin{table*}[tbh]
\begin{center}
\begin{tabularx}{\textwidth}{>{\centering\arraybackslash}c >{\centering\arraybackslash}c >{\centering\arraybackslash}c >{\centering\arraybackslash}c >{\centering\arraybackslash}c >{\centering\arraybackslash}c >{\centering\arraybackslash}X}
\hline \hline 
System & Source &  $c_{00}$  & $\kappa_{\alpha}$ & $\kappa_{r}$  &  $\kappa_{n}$ & $\kappa_{e}$  \\
\hline 
\multirow{2}{*}{Al$^{+}$/ Hg$^{+}$~\cite{AlHgdrift}} & Sun & $-3.0 \ (5.7) \times 10^{-7} $~\cite{Dzuba2018} & $5.3 \ (10) \times 10^{-8}$~\cite{Dzuba2018} & $-6.7 \ (13) \times 10^{-5}$ & $-3.1 \ (6.0) \times 10^{-3}$ & --- \\
& Moon & $-6.0 \ (12) \times 10^{-3}$ & $1.1 \ (2.1) \times 10^{-3}$ & -0.67 \ (1.3) & -63 \ (120) & --- \\
$\mathrm{Yb}^{+}/\mathrm{Cs}$~\cite{Lange2021}  & Sun & $4.2 \ (3.3) \times 10^{-8}$  & $14 \ (11) \times 10^{-9}$~\cite{Lange2021} & $4.0 \ (3.1) \times 10^{-5}$ & $7 \ (45) \times 10^{-8}$~\cite{Lange2021} & $-7 \ (45) \times 10^{-8}$~\cite{Lange2021} \\
$\mathrm{Yb}^{+}/\mathrm{Yb}^{+}$~\cite{Filzinger2023} & Sun &  $-7.4 \ (9.3) \times 10^{-9}$  & $-2.4 \ (3.0) \times 10^{-9}$\cite{Filzinger2023} & $7.0 \ (8.7) \times 10^{-6}$ & $2.1 \ (2.9) \times 10^{-4}$ & --- \\
\hline \hline
\end{tabularx}
\caption{Limits on the Standard Model Extension parameter $c_{00}$ and the parameters $\kappa_{\alpha}$, $\kappa_{r}$, $\kappa_{n}$ and $\kappa_{e}$ describing the dependence of the fine structure constant $\alpha$, the internucleon distance $r_0$ (derived from the nuclear charge radius $r_N$), the nucleon mass and the electron mass on the gravitational potential.}
%gravity related variation of the nuclear charge radius $\kappa_{r}$ (or internucleon distance $r_0$) and the nucleon mass $\kappa_{n}$. }
\label{Kappatable}
\end{center}
\end{table*}

\section{Constraints on the Yukawa-type interaction mediated by the scalar field  produced by massive bodies}

% interaction between hypothetical scalar particles and standard model particles

In this section, we consider the Yukawa-type interaction mediated by the scalar field  produced by massive bodies. Let us first provide an introduction to the phenomenology of this scalar field, following Ref.~\cite{Leefer2016}. 
%Light scalar fields appear naturally in modern cosmological models. 
A scalar field $\phi$ interacts with the standard model sector via the Yukawa-type Lagrangian:

\begin{align} \label{Yukawainteraction}
    \mathcal{L}_{\text{int}} = - \sum_{f} \frac{\phi}{\Lambda_{f}} m_{f} \Bar{f}f + \frac{\phi}{\Lambda_{\gamma}} \frac{F_{\mu \nu} F^{\mu \nu}}{4} \,.
\end{align}
Here, the first term represents the coupling to fermion fields $f$, with mass $m_{f}$ and $\Bar{f} = f^{\dagger} \gamma_{0}$, while the second term represents the coupling to the photon field. $\Lambda_{f}$ and $\Lambda_{\gamma}$ are new-physics energy scales which determine the strength of these couplings. Adding these interaction terms to the relevant terms in the standard model Lagrangian

\begin{align} \label{Lagrangian}
    \mathcal{L} \supset - \sum_{f} m_{f} \Bar{f}f - \frac{F_{\mu \nu} F^{\mu \nu}}{4 \alpha} \,,
\end{align}
we observe that we may present the effects of the interaction terms in the form of a variable fermion mass $m_{f}$ and electromagnetic fine-structure constant $\alpha$  (see e.g ~\cite{Stadnik2015})

\begin{align}
    m_{f} & \rightarrow m_{f} \left( 1 + \frac{\phi}{\Lambda_{f}} \right) \,, \label{mfvar} \\
    \alpha & \rightarrow \frac{\alpha}{1 - \phi/\Lambda_{\gamma}} \approx \alpha \left( 1 + \frac{\phi}{\Lambda_{\gamma}} \right) \,. \label{alphavar}
\end{align}
Adding the kinetic term to the interaction Lagrangian (\ref{Yukawainteraction}) gives the following equations of motion for the field  $\phi$:  

\begin{align}
    (\partial_{\mu} \partial_{\mu} + m_{\phi}^{2}) \phi = - \sum_{f} \frac{m_{f}}{\Lambda_{f}} m_{f} \Bar{f}f + \frac{1}{\Lambda_{\gamma}} \frac{F_{\mu \nu} F^{\mu \nu}}{4} \,,
\end{align}
where $m_{\phi}$ is the mass of the scalar particle. This implies that in the presence of the interaction (\ref{Yukawainteraction}), the standard model fermion and photon fields act as sources of the scalar field. Massive bodies such as stars or galaxies, which are composed of atoms, may act as these sources, producing a scalar field mediating Yukawa-type interactions. This field may produce a local variation of fundamental constants in the presence of a massive body with a varying distance to the laboratory. As such, we may investigate the influence of the scalar particle $\phi$ on atomic spectroscopy experiments, see Ref.~\cite{Leefer2016}. In the following subsections we place constraints on the scalar field's interaction with the Standard Model particles. Specifically, we consider the effects produced by a variation in the scalar field due to both the semi-annual variation in the Sun-Earth distance, and the approximately semi-monthly variation in the Moon-Earth distance. These constraints are compared to those obtained from tests of the equivalence principle by the MICROSCOPE mission~\cite{MICROSCOPE1,MICROSCOPE2}.

% we consider the measurement of the variation in the ratio of optical clock transitions in Al$^{+}$/Hg$^{+}$, and

\subsection{Effects produced by the variation in the Sun-Earth distance}

% 
% Let us now consider specifically the effects of the interactions between the scalar particles produced by the Sun. 

Similar to the gravitational potential, the scalar Yukawa potential  depends on the distance between Sun and Earth.
Assuming the Sun's elemental composition to be 75\% $^{1}$H and 25\% $^{4}$He by mass, the resultant scalar field  may be expressed as~\cite{Leefer2016}

\begin{widetext}

\begin{align} \label{Yukawa}
    \phi_{\text{Sun}} = - N_s m_{n}\left[ \frac{0.15}{\Lambda_{n^{\prime}}} + 1.1 \biggl\{  \frac{1}{\Lambda_{p}} + \frac{5 \times 10^{-4}}{\Lambda_{e}}   \biggr\}  + \frac{8 \times 10^{-4}}{\Lambda_{\gamma}}  \right] \frac{e^{- m_{\phi} r}}{4 \pi r} = -N_s \beta_{s} \frac{e^{- m_{\phi} r}}{4 \pi r}  \,,
\end{align}

\end{widetext}
where $m_{n} = (m_{p} + m_{n^{\prime}})/2 = 0.94$ GeV is the average nucleon mass and $N_s$  
%In general, we cannot treat an extended source of the scalar field $\phi_{s}$ as a point mass located at its origin, even in a system with spherical symmetry, unless the distance between the source body and the measurement apparatus is much greater than the dimensions of both. Thus, we define the total scalar field from the sun at position $ \bf{r}$ to be
%
%\begin{align} 
%    \Phi_{s,\text{Sun}}(\mathbf{r}) = \int_{V} n_{s}(\mathbf{r}^{\prime}) \phi_{s,\text{Sun}} (\mathbf{r} - \mathbf{r^{\prime}}) d^{3} \mathbf{r}^{\prime} = - \frac{\beta_{s}}{4 \pi} \int_{V} n_{s} (\mathbf{r}^{\prime}) \frac{e^{- m_{\phi} \lvert \mathbf{r} - \mathbf{r^{\prime}} \rvert}}{\lvert \mathbf{r} - \mathbf{r^{\prime}} \rvert} d^{3} \mathbf{r}^{\prime} \equiv  \beta_{s} \mathcal{F}_{s}(m_{\phi}, \mathbf{r}) \,, \label{Totalphisun}
%\end{align}
%\end{widetext}
%where $n_{s}$ is the number density of atoms in the Sun. If the Compton wave length $\hbar/m_{\phi}c$ of the scalar mediator exceeds the size of the Sun, the result of the integration is  close to the Yukawa field produced by the point-like source, 
%Eq. (\ref{Yukawa}) multiplied by
is the number of atoms inside the Sun. The number of nucleons inside the Sun may be determined as mass of the Sun divided by the proton mass, $N_s=M_s/m_p=1.99 \times  10^{30} \ \text{kg}/1.67 \times 10^{-27} \ \text{kg}=1.19 \times  10^{57}$. The ratio of the number of neutrons and protons depends on the composition of the Sun, which is mainly composed of hydrogen and helium. According to Eq. (\ref{Yukawa}), the average atom in the Sun contains 1.1 protons and 0.15 neutrons, i.e. 1.25 nucleons. This implies that the number of atoms in the Sun is $N_s=0.95 \times  10^{57}$. In obtaining limits on the nucleon constant $\Lambda_{n}$ we consider the sum of the proton and neutron contributions assuming $\Lambda_{n^{\prime}}=\Lambda_{p}$. In this case, the limits on $\Lambda_{n}$ have no dependence on the composition of the Sun. The Earth's orbit is elliptical, with the Earth-Sun distance changing between $1.52 \times 10^{8}$ km and $1.47 \times 10^{8}$ km.

Using new and existing calculations of the variation of fundamental constants due to changes in the Yukawa potential (see Appendix \ref{AppendixConstants}), we place constraints on the Yukawa-type interactions of the scalar field from the Sun with photons, nucleons and electrons for a range of scalar particle masses, using a similar method to Ref.~\cite{Leefer2016}. 
Constraints on the parameters  $\Lambda_{\gamma}, \Lambda_{n}$ and $\Lambda_{e}$ are presented in Figure \ref{FigSensitivity} and summarised in Table \ref{Tab}.

% In addition to this, we use the limits on the coupling of the electron to proton mass ratio $\mu = m_{e}/m_{p}$ to gravity presented in Ref.~\cite{Lange2021} to place constraints on the interaction of the scalar field from the Sun with electrons and nucleons.

\subsection{Effects produced by the variation in the Moon-Earth distance}

% In a similar way, we may also consider the scalar field produced by the moon. 

We may also extract limits on these parameters by considering the varying Yukawa potential of the Moon's orbit around Earth. Doing so allows one to investigate the coupling at larger values of the scalar particle mass $m_{\phi}$. The average atom in the Moon contains 12 protons and 12 neutrons. Following Ref.~\cite{Leefer2016} we obtain the Moon's scalar field to be
%elemental composition is a 1:1 ratio of number densities of $^{24}$M$^{16}$O and $^{28}$S$^{16}$O$^{2}$, the resultant scalar field may be expressed as~\cite{Leefer2016}

\begin{widetext}

\begin{align}\label{Moon}
    \phi_{\text{Moon}} = -N_m m_{n}\left[ \frac{12}{\Lambda_{n^{\prime}}} + 12 \biggl\{  \frac{1}{\Lambda_{p}} + \frac{5 \times 10^{-4}}{\Lambda_{e}}   \biggr\}  + \frac{0.03}{\Lambda_{\gamma}}  \right] \frac{e^{- m_{\phi} r}}{4 \pi r} = -N_m \beta_{m} \frac{e^{- m_{\phi} r}}{4 \pi r}  \,.
\end{align}
%As per Equation (\ref{Totalphisun}), we define the total scalar field at position $\bf{r}$ from the moon to be
%\begin{align} 
%    \Phi_{s,\text{Moon}}(\mathbf{r}) = \int_{V} n_{m}(\mathbf{r}^{\prime}) \phi_{s,\text{Moon}} (\mathbf{r} - \mathbf{r^{\prime}}) d^{3} \mathbf{r}^{\prime} = - \frac{\beta_{m}}{4 \pi} \int_{V} n_{m} (\mathbf{r}^{\prime}) \frac{e^{- m_{\phi} \lvert \mathbf{r} - \mathbf{r^{\prime}} \rvert}}{\lvert \mathbf{r} - \mathbf{r^{\prime}} \rvert} d^{3} \mathbf{r}^{\prime} \equiv \beta_{m} \mathcal{F}_{m}(m_{\phi}, \mathbf{r}) \,, \label{Totalphimoon}
%\end{align}

\end{widetext}
The number of nucleons in the Moon may be determined as the mass of the Moon divided by the proton mass, $N_m=M_m/m_p=  7.34 \times  10^{22} \ \text{kg}/1.67 \times 10^{-27} \ \text{kg}=4.40 \times  10^{49}$.  %According to Eq. (\ref{Yukawa}),
The average atom in the Moon contains  24 nucleons. This implies that the number of atoms in the Moon to be $N_m= 1.8 \times  10^{48}$. In obtaining limits on the nucleon constant $\Lambda_{n}$ we once again consider the sum of the proton and neutron contributions assuming $\Lambda_{n^{\prime}}=\Lambda_{p}$.  In this case, the limits on $\Lambda_{n}$ have no dependence on the composition of the Moon.

%Once again, if the scalar particle's Compton wave length $\hbar/m_{\phi}c$ is bigger than the distance to the source, the scalar field $\Phi_{s}$ has a $1/r$ dependence (for a spherically symmetric source), that is proportional to the local gravitational potential $U = -GM/r$. As such, we seek to calculate limits on the photon and nucleon coupling constants, $\Lambda_{\gamma}$ and $\Lambda_{n}$ via an analysis of the correlation between $\alpha$, $m_{N}$ and the varying gravitational potential in the laboratory due to the Moon's eccentric orbit around the Earth. 

The average Earth-Moon distance is $3.84 \times 10^{5} $ km, centre to centre, and this value varies between $3.69 \times 10^{5}$ km and $3.99 \times 10^{5}$ km with a period of approximately 27.3 days.  Furthermore, due to the relatively large diameter of the Earth ($\sim 1.27 \times 10^{4}$ km), we note that there is also a daily variation in the distance between the Moon and the laboratory. Despite this, the contributions from the monthly variations are more significant, and as such we consider a variation of $\sim 3.0 \times 10^{4}$ km in the Earth-Moon distance, which corresponds to the minimal seasonal variation (see Ref.~\cite{Leefer2016}). 

% This variation once again provides us with a means of investigating the influence of the scalar particle $\phi_{s}$ on atomic spectroscopy experiments. 

\begin{figure*}[tbh]
\begin{tabular}{cc}
	\includegraphics[width=8.5cm]{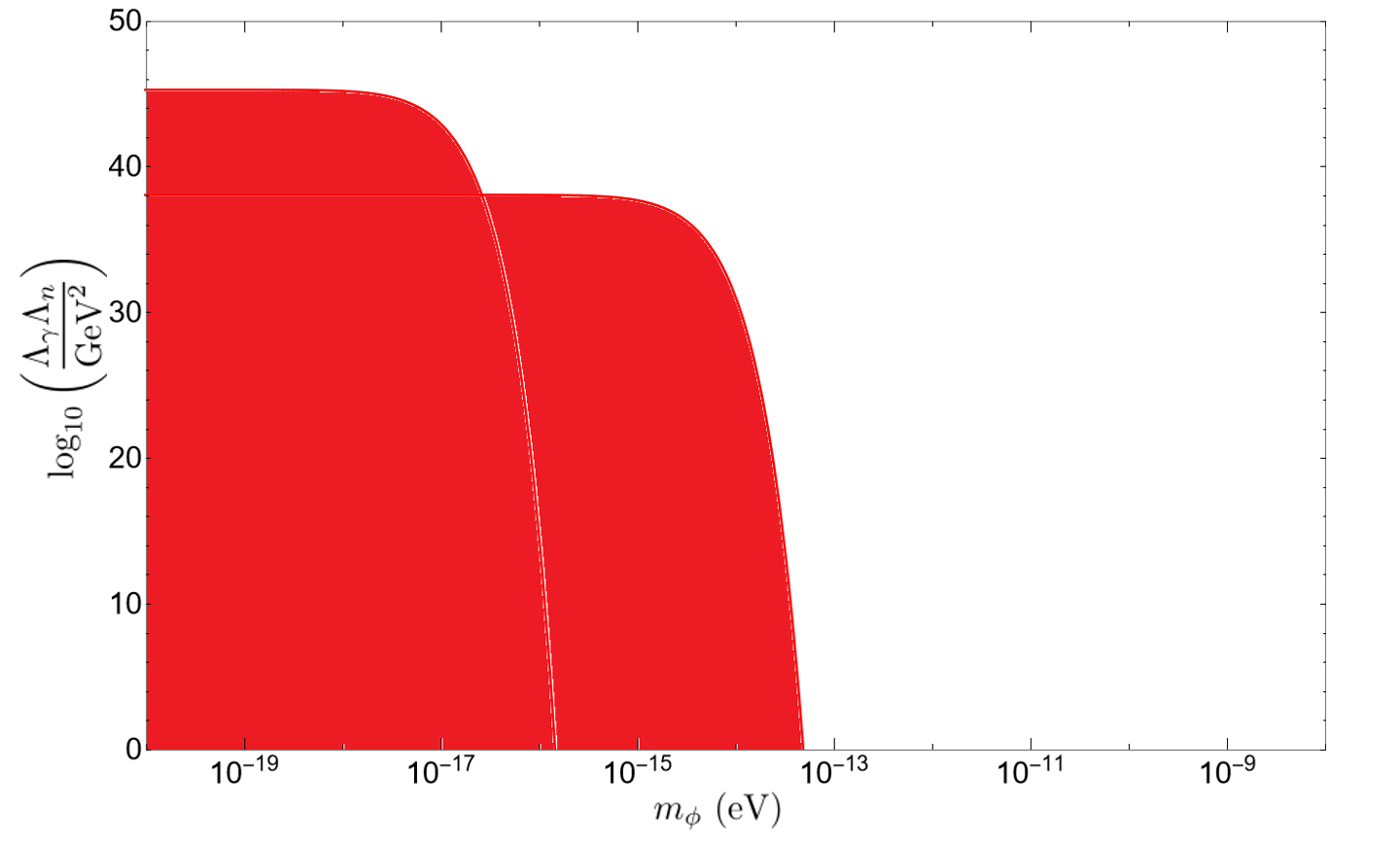} & 
	\includegraphics[width=8.5cm]{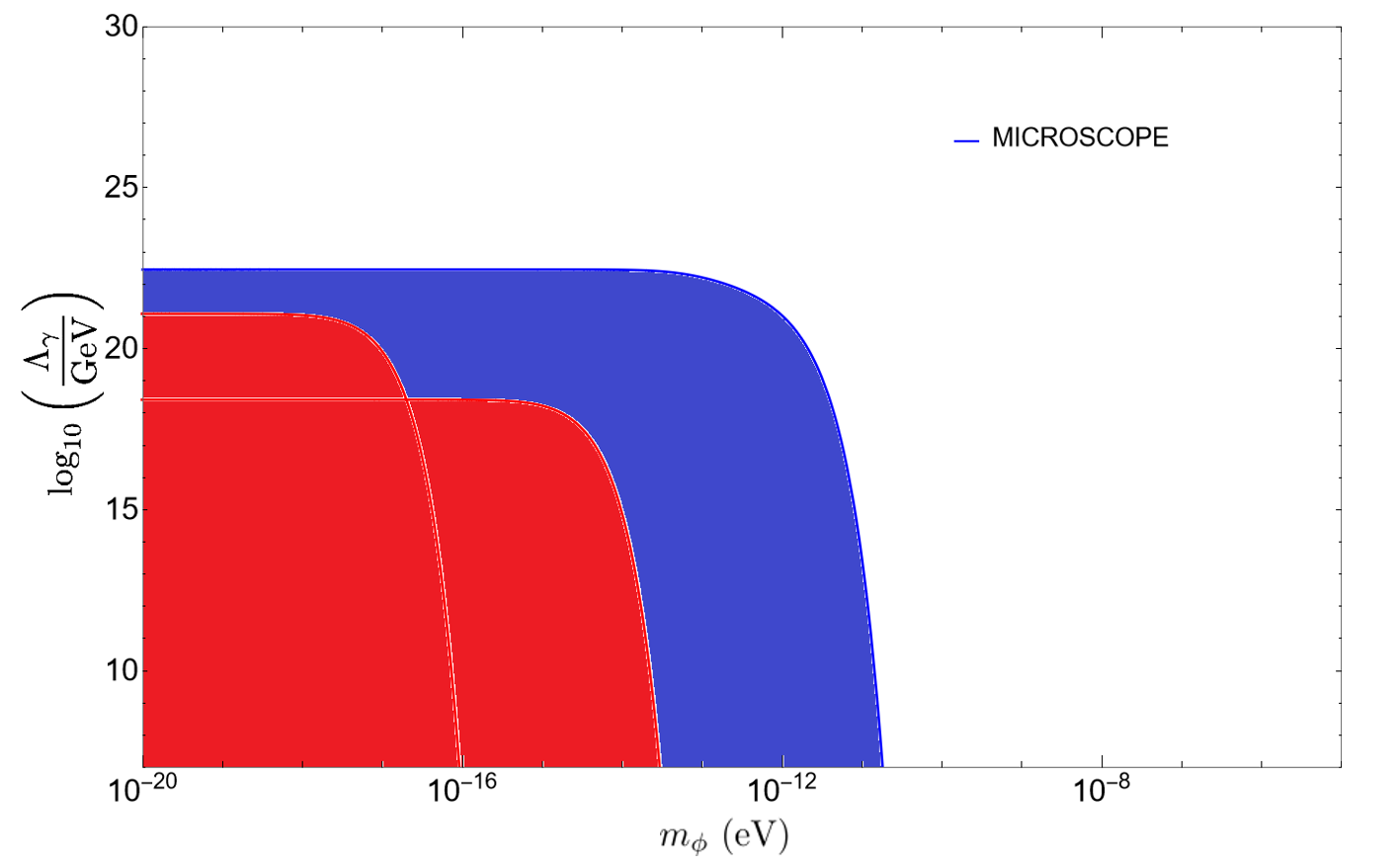} \\
    \includegraphics[width=8.5cm]{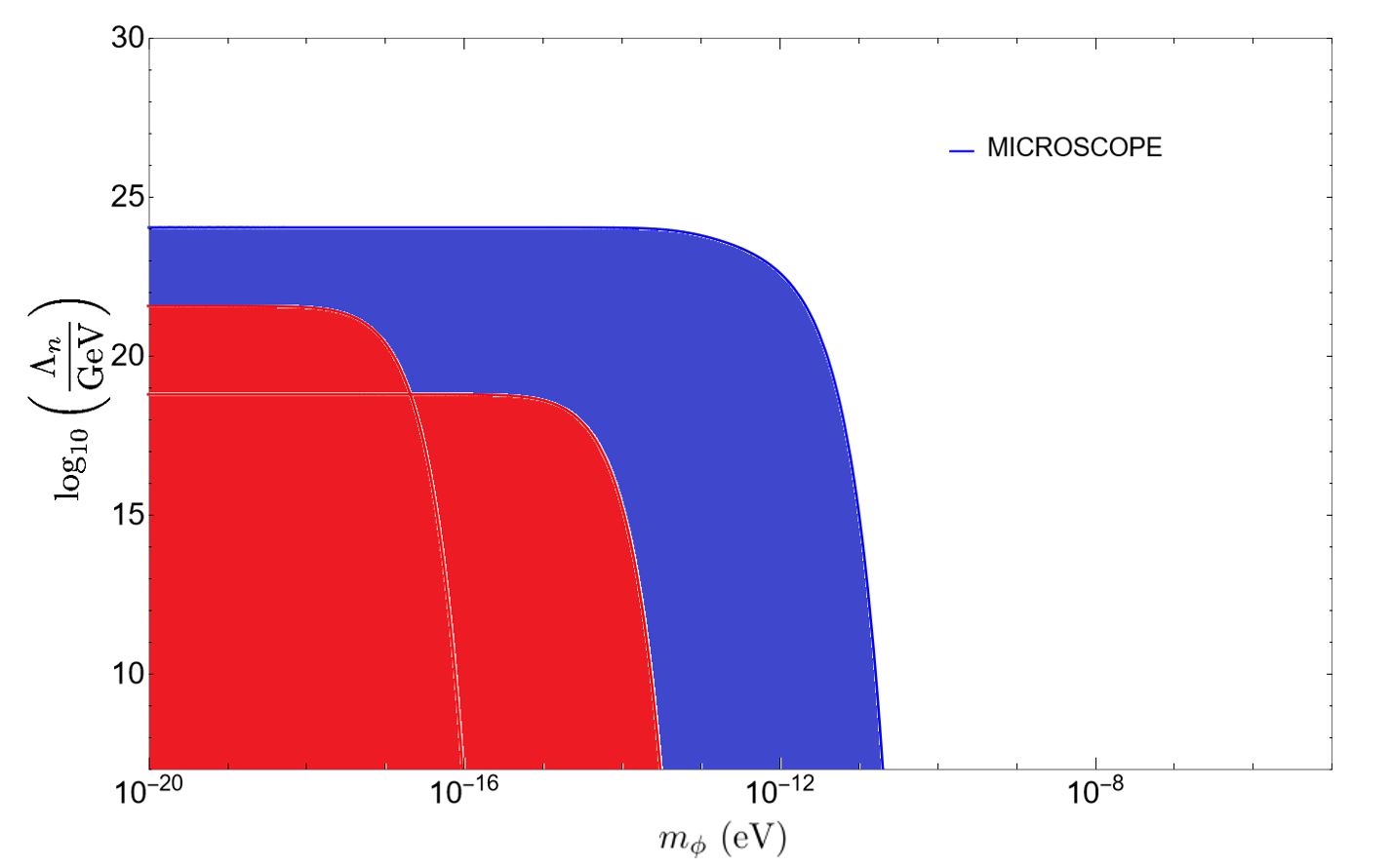} &
    \includegraphics[width=8.5cm]{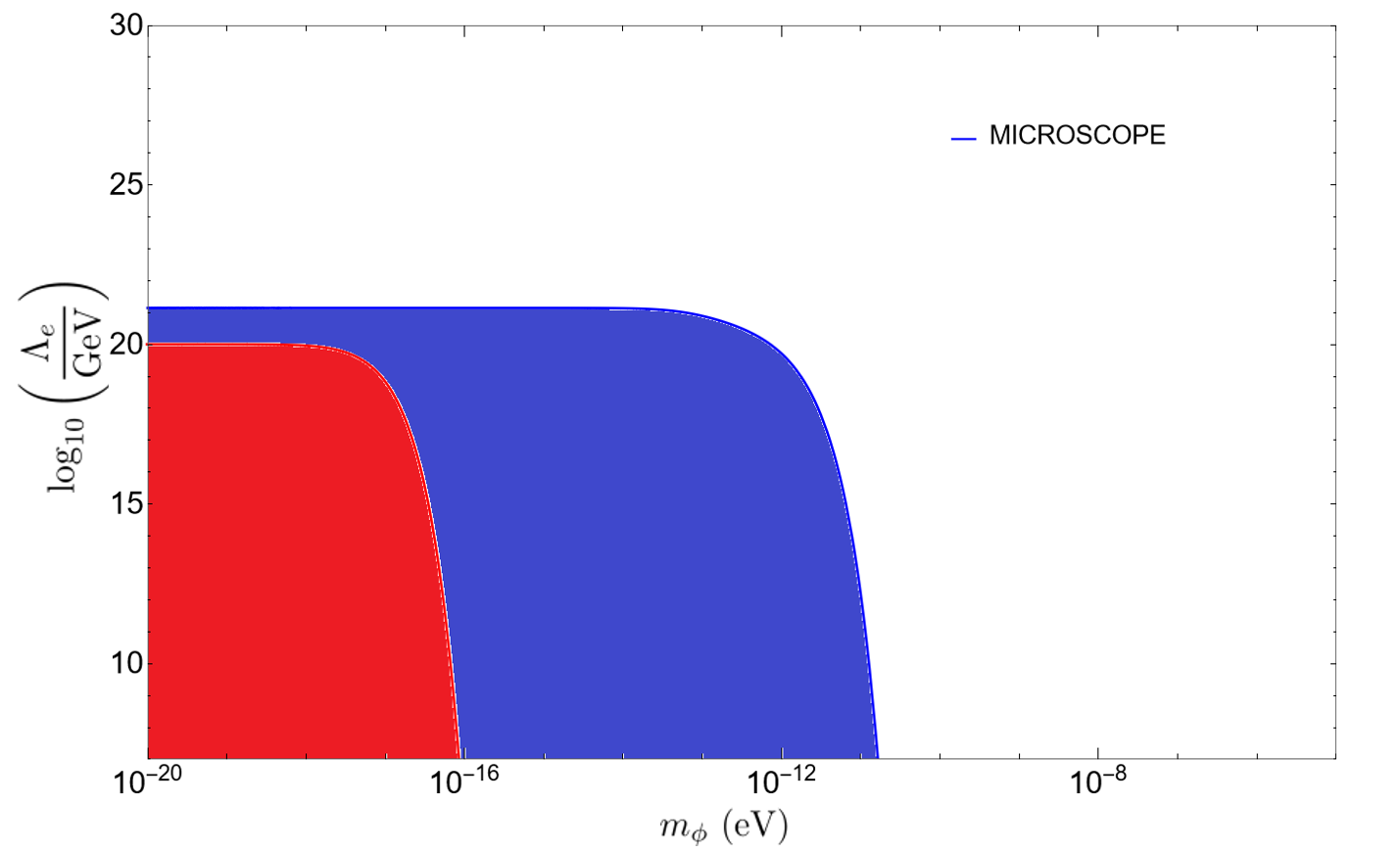} 
\end{tabular}
	\caption{Limits on the constants of the Yukawa-type interaction Eq. (\ref{Yukawainteraction}) of the scalar field $\phi_{s}$ with photons ($\Lambda_{\gamma}$), nucleons ($\Lambda_{n}$) and electrons ($\Lambda_{e}$). In obtaining limits on the nucleon constant $\Lambda_{n}$ we sum the contributions from the proton and neutron coupling constants, assuming $\Lambda_{n^{\prime}}=\Lambda_{p}$. The upper curve depicts the region constrained by the  potential of the Sun, whilst the lower and longer curve depicts the region constrained by the Moon's  potential. There is no data for the effect of the Moon's gravitational potential on the electron interaction constant (lower right tile). The region in blue shows the parameter space which is excluded by the MICROSCOPE experiment~\cite{MICROSCOPE1,MICROSCOPE2}.}
    \label{FigSensitivity}
\end{figure*}

% The Upper curve shows the region constrained exclusion plot obtained from the  Sun potential, lower and longer curve shows the exclusion plot for the Moon potential. There are no Moon data for the electron interaction constant (last graph).

%Using a similar method to above, we seek to approximate changes in the Moon's gravitational potential on earth by using the relation $\delta U_{\text{moon}} = G M_{\odot} \epsilon/a_{m}$ to first order in the eccentricity of Moon's orbit, $\epsilon = 0.0549$. Here, $a_{m} = 0.002$ corresponds to the semi-major axis of the Moon's orbit around Earth, which yields a semi-annual change in the laboratory gravitational potential of $(2\delta U_{\text{moon}})/c^{2} \approx 1.62 \times 10^{-14}$, corresponding to a change in the Earth-Moon distance between $3.69 \times 10^{5}$ km and $3.99 \times 10^{5}$ km. 

% The amplitude of the variation in the Earth-Moon distance varies by $\sim \pm 10^{4}$ km 

Thus, using a similar method to that of the previous subsection, we may determine limits on the Yukawa-type interaction between the varying scalar field  and photons/nucleons for a range of scalar particle masses. Constraints on these parameters are presented in Fig.~\ref{FigSensitivity}, and summarised in Table \ref{Tab}. 

These limits may be compared to the limits from the atomic spectroscopy measurements presented in Refs.~\cite{Leefer2016,Brzeminski2022}. Our limits based on the Sun and Moon data are significantly stronger than that of those presented in Ref.~\cite{Leefer2016}. Despite this, their results also contain limits on the scalar field produced by a 300 kg lead mass on a distance of 1 m. This area is sensitive to much bigger scalar masses and is not covered by our results. Further, our constraint on the photon coupling constant $\Lambda_{\gamma}$ is stronger than the corresponding limit obtained from existing atomic clock experiments presented in Figure 5 of Ref.~\cite{Brzeminski2022}. This is due to the fact that the constraint presented in this reference uses the results obtained from Yb$^{+}$/Cs experiments~\cite{Lange2021} only, while our constraints are based on the more sensitive Yb$^{+}$/Yb$^{+}$ experiments~\cite{Filzinger2023}.

Constraints yielded from the results of existing atomic clock experiments do not exceed the sensitivity provided by direct fifth force searches such as the MICROSCOPE mission~\cite{MICROSCOPE1,MICROSCOPE2} (see Figure \ref{FigSensitivity}). In order for them to do so, we require an improvement of $\sim 3$ orders of magnitude in the fractional frequency uncertainty in optical clocks located on Earth, something which is predicted to occur over the next few decades~\cite{Derevianko2021}. Alternatively, there exist proposals for experiments with clocks based on nuclear transitions~\cite{Peik2020}, which are very  sensitive to the variation of fundamental constants \cite{Flambaum2006},  and optical clock experiments conducted in space~\cite{FOCOS,SpaceQ}. Constraints yielded from such experiments are predicted to match/exceed the sensitivity of direct fifth force searches~\cite{Brzeminski2022}.

\subsection{Limits on combinations of coupling constants}
Spectroscopy measurements may also be used to probe different combinations of the couplings constants $\Lambda_{\gamma,p,n,e}$~\cite{Leefer2016}, some of which may not be otherwise probed using anomalous-force measurements. Such results  may be useful in the case when the variation of the ratio of optical clock transition frequencies and the source-dependent functions $\beta_{s,m}$ are dominated by different terms. Thus we may use the results of the previous two subsections to provide constraints on the combination of parameters $\Lambda_{\gamma}\Lambda_{n}$, see Figure \ref{FigSensitivity}. 
 If the scalar particle's Compton wave length $\hbar/m_{\phi}c$ is bigger than the distance to the source,
%in the limit that the scalar particle mass $m_{\phi} \rightarrow 0$, 
we determine the following limits for the combination of parameters $\beta_{s,n}/\Lambda_{\gamma}$ 
\begin{align}
    \frac{\Lambda_{\gamma}}{\beta_{s}} & \gtrsim 2 \times 10^{45} \ \text{GeV}^{2}\,,  \\
    \frac{\Lambda_{\gamma}}{\beta_{m}} & \gtrsim 1 \times 10^{38} \ \text{GeV}^{2}\,.
\end{align}
%These limits are 1-2 orders of magnitude stronger than the corresponding limits presented in Ref.~\cite{Leefer2016}.

% \begin{figure*}[tbh]
% 	\includegraphics[width=12cm]{CrossTerm.png} 
% 	\caption{Limits on the Yukawa-type interaction of the scalar field $\phi_{s}$ on the combination of parameters $\Lambda_{\gamma} \Lambda_{n}$.}
%     \label{CrossTermFigSensitivity}
% \end{figure*}

\renewcommand{\arraystretch}{2}
\begin{table*}
\begin{center}
\begin{tabularx}{\textwidth}{ >{\centering\arraybackslash}X| >{\centering\arraybackslash}X| >{\centering\arraybackslash}X| >{\centering\arraybackslash}X| >{\centering\arraybackslash}X| >{\centering\arraybackslash}X}
\hline \hline
\multirow{2}{*}{System} & \multirow{2}{*}{Source/Attractor} & \multirow{2}{*}{$\Lambda_{\gamma}/\beta$} & \multirow{2}{*}{$\Lambda_{\gamma}$}  & \multirow{2}{*}{$\Lambda_{n}$} & \multirow{2}{*}{$\Lambda_{e}$} \\
 &  & (GeV$^2$) & (GeV) & (GeV) & (GeV)
\\ \hline \hline
\multirow{2}{*}{Al$^{+}$/ Hg$^{+}$~\cite{AlHgdrift}} & Sun & $7 \times 10^{43}$ & $2 \times 10^{20}$ & $1 \times 10^{21}$ & --- \\
 & Moon & $1 \times 10^{38}$ & $2 \times 10^{18}$ & $6  \times 10^{18}$ & --- \\
 \hline
Yb$^{+}$/ Cs~\cite{Lange2021} & Sun & $4 \times 10^{44}$ & $6 \times 10^{20}$  & $5 \times 10^{21}$ & $1 \times 10^{20}$ \\
\hline
Yb$^{+}$/ Yb$^{+}$~\cite{Filzinger2023} & Sun & $2 \times 10^{45}$ & $1 \times 10^{21} $ & $4 \times 10^{21}$ & --- \\
\hline \hline
\end{tabularx}
\caption{Lower limits on the constants of the Yukawa-type interaction Eq. (\ref{Yukawainteraction}) of the scalar field $\phi_{s}$ with photons ($\Lambda_{\gamma}$), nucleons ($\Lambda_{n}$) and electrons ($\Lambda_{e}$) in the interval of small scalar field masses determined by the condition %Summary of the obtained limits on the Yukawa-type interaction of the scalar field $\phi_{s}$ with photons, nucleons and electrons, in the case where 
that the scalar particle's Compton wave length $\hbar/m_{\phi}c$ is bigger than the distance to the source, $m_{\phi}  < 1.0 \times 10^{-18}$ eV for the Sun-Earth distance and $m_{\phi}  < 0.5 \times 10^{-15}$ eV for the Moon-Earth distance. In obtaining limits on the nucleon constant $\Lambda_{n}$ we sum the contributions from the proton and neutron coupling constants, assuming $\Lambda_{n^{\prime}}=\Lambda_{p}$. The full exclusion plots are presented in Fig. \ref{FigSensitivity}.}
\label{Tab}
\end{center}
\end{table*}

Limits on the variation of $\alpha$, the nucleon mass and the electron mass due to the variation of the Sun's distance to Earth are presented in Table \ref{GravCouplingTable}.
The details are presented in Appendix \ref{AppendixConstants}.

% due to variation of the distance to Sun

\begin{table*}[!tbh]
\begin{center}
\begin{tabularx}{\textwidth}{  >{\centering\arraybackslash}X >{\centering\arraybackslash}X >{\centering\arraybackslash}X >{\centering\arraybackslash}X }
\hline \hline 
System   &  $\delta \alpha/\alpha$  & $\delta m_{n}/m_{n}$ & $\delta m_{e}/m_{e}$ \\
\hline 
Al$^{+}$/ Hg$^{+}$~\cite{AlHgdrift} &  $0.17 \ (0.33) \times 10^{-16}$ & $1.2 \ (2.3) \times 10^{-15}$ & --- \\
$\mathrm{Yb}^{+}$/Cs~\cite{Lange2021} & $4.6 \ (3.6) \times 10^{-18}$ & $2.3 \ (15) \times 10^{-17}$ & $-2.3 \ (15) \times 10^{-17}$  \\
$\mathrm{Yb}^{+}/\mathrm{Yb}^{+}$~\cite{Filzinger2023} & $-7.9 \ (9.9) \times 10^{-19}$ & $1.3 \ (1.6) \times 10^{-16}$ & ---  \\

\hline \hline
\end{tabularx}
\caption{Summary of the obtained constraints on the variation of fundamental constants due to changes in the distance to the Sun (interpreted in Refs.~\cite{AlHgdrift, Lange2021,Filzinger2023} as the effect of the variation in the Sun's gravitational potential).}
\label{GravCouplingTable}
\end{center}
\end{table*}

% we sum proton and neutron contributions assuming $\Lambda_{n^{\prime}}=\Lambda_{p}$

% These limits are in agreement with the limits presented in Figure 2 of Ref.~\cite{Leefer2016}. Our results for the nucleon coupling constant $\Lambda_{n}$ predicts a larger probable scalar particle mass region than the limits presented in Ref.~\cite{Leefer2016}. 

\section{Summary}
% One of the most effective ways to search for signatures of ultralight dark matter's interaction with standard model particles is via atomic spectroscopy.
Atomic spectroscopy measurements are used to search for the potential space-time variation of physical constants.
%This variation may be caused by an interaction with a slowly evolving scalar dark energy field or an oscillating dark matter field}. 
In particular, we relate the proton mass $m_{p}$ and the quark mass $m_{q}$ variation to measurements in the variation of two frequency ratios: the comparison of the two-photon transition in atomic Hydrogen to the results from clocks based on $^{133}$Cs in Ref.~\cite{Fischer2004}, and the variation in the ratio of the two optical clock frequencies in Al$^{+}$ and Hg$^{+}$ in Ref.~\cite{AlHgdrift}. We used this data to place new limits on the variation of the proton mass $m_{p}$, as well as independent limits on the variation of the quark mass $m_{q}$, both of which may be used to place limits on the variation of the QCD parameter $\theta$.

In the second part of this paper we considered the beyond-standard-model effects of gravity, such as the Einstein Equivalence Principle (EEP) violating term ($c_{00}$) in the Standard Model Extension (SME) Lagrangian \cite{Kostelecky} and the dependence of fundamental constants on the gravitational potential, based on the measurements of the dependence of the ratio of atomic transition frequencies Al$^{+}$/Hg$^{+}$ ~\cite{Dzuba2018,AlHgdrift}, Yb$^{+}$/ Cs~\cite{Lange2021} and Yb$^{+}$/Yb$^{+}$~\cite{Filzinger2023} to the Sun-Earth distance.

In the third part of this paper we considered the scalar field produced by massive bodies. %Such particles appear naturally in modern cosmological models, and are theorised to have feeble interactions with Standard Model particles.
We determine limits on the interactions of the scalar particle with photons, nucleons and electrons for a wide range of scalar particle masses, 
basing on the measurements of dependence of the ratio of  atomic transition frequencies Al$^{+}$/Hg$^{+}$ ~\cite{Dzuba2018,AlHgdrift}, Yb$^{+}$/ Cs~\cite{Lange2021} and Yb$^{+}$/Yb$^{+}$~\cite{Filzinger2023} on the Sun-Earth and Moon-Earth distances.
%atomic spectroscopy data of the variation in the ratio of the two optical clock frequencies in Al$^{+}$ and Hg$^{+}$ presented in Refs.~\cite{Dzuba2018,AlHgdrift}, along with existing experimental limits on the , from Yb$^{+}$/ Cs~\cite{Lange2021} and Yb$^{+}$/Yb$^{+}$~\cite{Filzinger2023} systems.
% pre-existing theoretical calculations from Refs.~\cite{Flambaum2007,Dzuba2018} we determine limits on the 
%gravity related 
%variation of the nuclear radius and nuclear mass.These results are used to constrain the Yukawa type interactions of the scalar particle with photons, nucleons and electrons for a wide range of scalar particle masses. 
If the scalar particle's Compton wave length $\hbar/m_{\phi}c$ is bigger than the distance to the source,
we place the following limits on the coupling constants of the interaction of the scalar field with photons, nucleons and electrons: $\Lambda_{\gamma} \gtrsim 1 \times 10^{21} \ \text{GeV}$, $\Lambda_{n} \gtrsim 5 \times 10^{21} \ \text{GeV}$ and $\Lambda_{e} \gtrsim 1 \times 10^{20} \ \text{GeV}$.

% These limits may be compared to the limits from the atomic spectroscopy measurements presented in Ref.~\cite{Leefer2016}. Our limits based on the Sun and Moon data are significantly stronger. However, Ref.~\cite{Leefer2016} also obtained limits on the scalar field produced by a 300 kg lead mass on a distance of 1 m. This area is sensitive to much bigger scalar masses and is not covered by our results.

\section*{Acknowledgements}

The work was supported by the Australian Research Council Grants No.\ DP230101058 and No.\ DP200100150.

\appendix

\section{Coupling of fundamental constants to changes in the Yukawa  potential from the Sun/Moon} \label{AppendixConstants}

In this appendix we detail the calculations of the variation of fundamental constants due to changes in the Yukawa  potential. We will use experimental data on the variation of atomic transition frequencies as functions of Sun-Earth and Moon-Earth  distances motivated by the search for dependence of the fundamental constants on the gravitational potential.  Constraints on this dependence %presented in Table \ref{GravCouplingTable} and
were  obtained for all the systems of interest: Al$^{+}$/Hg$^{+}$, Yb$^{+}$/Cs and Yb$^{+}$/Yb$^{+}$. The  summary is presented in Table \ref{GravCouplingTable}.

Let us start with the variation of the fine structure constant $\alpha$. % Using Equation (\ref{alphagravity}) we may substitute the corresponding values of $\kappa_{\alpha}$ presented in Table \ref{Kappatable} to place numerical constraints on the variation of $\alpha$ due to a change in the gravitational potential. Noting that 
Equation (\ref{alphavar}) implies the following relation

\begin{align}
    \frac{\delta \alpha}{\alpha} = \frac{\delta \phi}{\Lambda_{\gamma}}\,,
\end{align}
At $m_{\phi} \rightarrow 0$ we have from (\ref{Yukawa}) or (\ref{Moon}) 
\begin{align}
   \delta \phi = -\frac{\beta N_{S,M}}{4\pi}\delta\left(\frac{1}{r}\right)_{S,M} \equiv \beta D_{S,M}.
\end{align}
Here  $N_{S,M}$ is number of atoms in the Sun or the Moon ($N_S \approx 0.95 \times 10^{57}$ and $N_M \approx 1.8  \times 10^{48}$) 
%$N_M \approx 9 \times 10^{47}$)
and $\delta (1/r)_{S,M}$ is half-yearly variation of the inverted Sun-Earth or Moon-Earth distance.
Then
\begin{align}
  \left|\Lambda_{\gamma}/\beta\right| = \left|\frac{D_{S,M}}{\delta \alpha/\alpha}\right|,
\end{align}
and
\begin{align}\label{e:lambda}
  \Lambda_{\gamma} = \sqrt{\left|\frac{m_n a_{\gamma}D_{S,M}}{\delta \alpha/\alpha}\right|},
\end{align}
where $a_{\gamma} = 8 \times 10^4$ for the Sun, and $a_{\gamma} = 0.03$ for the Moon (see  (\ref{Yukawa}) and (\ref{Moon})) .

Now let us consider the variation of the fermionic masses, of which we consider the nucleon and electron mass, $m_{n}$ and $m_{e}$ respectively. 
%We calculate the variation of the nucleon mass $m_{n}$ due to changes in the gravitational potential for systems Al$^{+}$/Hg$^{+}$ and Yb$^{+}$/Yb$^{+}$ using Equation (\ref{AlHgVariationRatio}), noting the fact that 
From Eq. (\ref{rpi}), we have $\frac{\delta r_{0}}{r_{0}}=1.2 \frac{\delta m_{\pi}}{m_{\pi}}$, where $r_0$ is the internucleon distance and $m_{\pi}$ is the pion mass. From Eq. (\ref{pionproton}) we see that $\frac{\delta m_{p}}{m_{p}} = 0.13 \frac{\delta m_{\pi}}{m_{\pi}}$ 
, where $m_p$ is the proton mass. Upon substitution into Eq. (\ref{rpi}) , we yield 

\begin{align}
  %  \frac{\delta r_{0}}{r_{0}} = 9.23 \frac{\delta m_{n}}{m_{n}}\,.
  \frac{\delta m_{n}}{m_{n}}= 0.11 \frac{\delta r_{0}}{r_{0}}.
\end{align}
This relation  indicates that the nucleon mass is less sensitive to variations in the pion mass (or quark mass) compared to the internucleon distance $r_0$. This is consistent with the fact that the quark contribution to nucleon mass is relatively small, while $r_0$  is very sensitive to the pion exchange potential. 
Thus, we yield

\begin{align} \label{e:mm}
   \frac{\delta m_{n}}{m_{n}} = \frac{\delta (\nu_{a}/\nu_{b})}{(\nu_{a}/\nu_{b})} \left( \frac{1}{18.5 (K_1-K_2) + (K_3-K_4} \right) \,, 
\end{align}
where $K_1$, $K_2$, $K_3$   and $K_4$ are defined in Equation (\ref{variation}). 
For the Al$^{+}$/Hg$^{+}$ and Yb$^{+}$/Yb$^{+}$ systems we use the calculated values of the field shift and mass shift constants (Ref.~\cite{Hur} for Yb$^+$ and the present work for Hg$^+$) 
and the experimental limits on the variation of the frequency ratios~\cite{AlHgdrift,Filzinger2023} to obtain the mass variation from (\ref{e:mm}).
For the Yb$^{+}$/Cs system, we use the limits on the coupling of the electron to proton mass ratio $\mu = m_{p}/m_{e}$ to gravity presented in Ref.~\cite{Lange2021} to place constraints on the fractional variation of the nucleon and electron mass due to changes in the distance to Sun. These calculations may once again be used to place constraints on the Yukawa-type interactions of the scalar field from the Sun/Moon with nucleons and electrons, noting the following relations which may be yielded from Equation (\ref{mfvar})

\begin{align}
    \frac{\delta m_{n}}{m_{n}} = \frac{\delta \phi_{s}}{\Lambda_{n}} \,, \\
    \frac{\delta m_{e}}{m_{e}} = \frac{\delta \phi_{s}}{\Lambda_{e}} \,.
\end{align}
Then following the same steps as for $\Lambda_{\gamma}$ we obtain
\begin{align}\label{e:lambdan}
  \Lambda_n = \sqrt{\left|\frac{m_n a_n D_{S,M}}{\delta m_n/m_n}\right|},
\end{align}
where $a_n=1.25 m_n$ for the Sun and $a_n=24 m_n$ for the Moon.
To get an expression for $\Lambda_e$ we replace $a_n$ by $a_e$ and $\delta m_n/m_n$ by $\delta m_e/m_e$ in (\ref{e:lambdan});
$a_n=5.5\times 10^{-4} m_n$ for the Sun and $a_n=6\times 10^{-3} m_n$ for the Moon.

Note that in the case of the nucleon coupling constant $\Lambda_{n}$ it is also possible to perform calculations using the values of $\kappa_{n}$, presented in Table \ref{Kappatable}, however in this case the results provide weaker constraints.

\bibliographystyle{apsrev4-2}
\bibliography{References.bib}

\begin{thebibliography}{67}%
\makeatletter
\providecommand \@ifxundefined [1]{%
 \@ifx{#1\undefined}
}%
\providecommand \@ifnum [1]{%
 \ifnum #1\expandafter \@firstoftwo
 \else \expandafter \@secondoftwo
 \fi
}%
\providecommand \@ifx [1]{%
 \ifx #1\expandafter \@firstoftwo
 \else \expandafter \@secondoftwo
 \fi
}%
\providecommand \natexlab [1]{#1}%
\providecommand \enquote  [1]{``#1''}%
\providecommand \bibnamefont  [1]{#1}%
\providecommand \bibfnamefont [1]{#1}%
\providecommand \citenamefont [1]{#1}%
\providecommand \href@noop [0]{\@secondoftwo}%
\providecommand \href [0]{\begingroup \@sanitize@url \@href}%
\providecommand \@href[1]{\@@startlink{#1}\@@href}%
\providecommand \@@href[1]{\endgroup#1\@@endlink}%
\providecommand \@sanitize@url [0]{\catcode `\\12\catcode `\$12\catcode `\&12\catcode `\#12\catcode `\^12\catcode `\_12\catcode `\%12\relax}%
\providecommand \@@startlink[1]{}%
\providecommand \@@endlink[0]{}%
\providecommand \url  [0]{\begingroup\@sanitize@url \@url }%
\providecommand \@url [1]{\endgroup\@href {#1}{\urlprefix }}%
\providecommand \urlprefix  [0]{URL }%
\providecommand \Eprint [0]{\href }%
\providecommand \doibase [0]{https://doi.org/}%
\providecommand \selectlanguage [0]{\@gobble}%
\providecommand \bibinfo  [0]{\@secondoftwo}%
\providecommand \bibfield  [0]{\@secondoftwo}%
\providecommand \translation [1]{[#1]}%
\providecommand \BibitemOpen [0]{}%
\providecommand \bibitemStop [0]{}%
\providecommand \bibitemNoStop [0]{.\EOS\space}%
\providecommand \EOS [0]{\spacefactor3000\relax}%
\providecommand \BibitemShut  [1]{\csname bibitem#1\endcsname}%
\let\auto@bib@innerbib\@empty
%</preamble>
\bibitem [{\citenamefont {Uzan}(2011)}]{Uzan}%
  \BibitemOpen
  \bibfield  {author} {\bibinfo {author} {\bibfnamefont {J.~P.}\ \bibnamefont {Uzan}},\ }\href {https://doi.org/https://doi.org/10.12942/lrr-2011-2} {\bibfield  {journal} {\bibinfo  {journal} {Living Reviews in Relativity}\ }\textbf {\bibinfo {volume} {14}},\ \bibinfo {pages} {2} (\bibinfo {year} {2011})}\BibitemShut {NoStop}%
\bibitem [{\citenamefont {Webb}\ \emph {et~al.}(2011)\citenamefont {Webb}, \citenamefont {King}, \citenamefont {Murphy}, \citenamefont {Flambaum}, \citenamefont {Carswell},\ and\ \citenamefont {Bainbridge}}]{WebbPRL2011}%
  \BibitemOpen
  \bibfield  {author} {\bibinfo {author} {\bibfnamefont {J.~K.}\ \bibnamefont {Webb}}, \bibinfo {author} {\bibfnamefont {J.~A.}\ \bibnamefont {King}}, \bibinfo {author} {\bibfnamefont {M.~T.}\ \bibnamefont {Murphy}}, \bibinfo {author} {\bibfnamefont {V.~V.}\ \bibnamefont {Flambaum}}, \bibinfo {author} {\bibfnamefont {R.~F.}\ \bibnamefont {Carswell}},\ and\ \bibinfo {author} {\bibfnamefont {M.~B.}\ \bibnamefont {Bainbridge}},\ }\href {https://doi.org/10.1103/PhysRevLett.107.191101} {\bibfield  {journal} {\bibinfo  {journal} {Phys. Rev. Lett.}\ }\textbf {\bibinfo {volume} {107}},\ \bibinfo {pages} {191101} (\bibinfo {year} {2011})}\BibitemShut {NoStop}%
\bibitem [{\citenamefont {King}\ \emph {et~al.}(2012)\citenamefont {King}, \citenamefont {Webb}, \citenamefont {Murphy}, \citenamefont {Flambaum}, \citenamefont {Carswell}, \citenamefont {Bainbridge}, \citenamefont {Wilczynska},\ and\ \citenamefont {Koch}}]{MNRAS2011}%
  \BibitemOpen
  \bibfield  {author} {\bibinfo {author} {\bibfnamefont {J.~A.}\ \bibnamefont {King}}, \bibinfo {author} {\bibfnamefont {J.~K.}\ \bibnamefont {Webb}}, \bibinfo {author} {\bibfnamefont {M.~T.}\ \bibnamefont {Murphy}}, \bibinfo {author} {\bibfnamefont {V.~V.}\ \bibnamefont {Flambaum}}, \bibinfo {author} {\bibfnamefont {R.~F.}\ \bibnamefont {Carswell}}, \bibinfo {author} {\bibfnamefont {M.~B.}\ \bibnamefont {Bainbridge}}, \bibinfo {author} {\bibfnamefont {M.~R.}\ \bibnamefont {Wilczynska}},\ and\ \bibinfo {author} {\bibfnamefont {F.~E.}\ \bibnamefont {Koch}},\ }\href {https://doi.org/10.1111/j.1365-2966.2012.20852.x} {\bibfield  {journal} {\bibinfo  {journal} {Monthly Notices of the Royal Astronomical Society}\ }\textbf {\bibinfo {volume} {422}},\ \bibinfo {pages} {3370} (\bibinfo {year} {2012})}\BibitemShut {NoStop}%
\bibitem [{\citenamefont {FLAMBAUM}\ and\ \citenamefont {BERENGUT}(2009)}]{Berengut}%
  \BibitemOpen
  \bibfield  {author} {\bibinfo {author} {\bibfnamefont {V.~V.}\ \bibnamefont {FLAMBAUM}}\ and\ \bibinfo {author} {\bibfnamefont {J.~C.}\ \bibnamefont {BERENGUT}},\ }\href {https://doi.org/10.1142/S0217751X0904693X} {\bibfield  {journal} {\bibinfo  {journal} {International Journal of Modern Physics A}\ }\textbf {\bibinfo {volume} {24}},\ \bibinfo {pages} {3342} (\bibinfo {year} {2009})},\ \Eprint {https://arxiv.org/abs/https://doi.org/10.1142/S0217751X0904693X} {https://doi.org/10.1142/S0217751X0904693X} \BibitemShut {NoStop}%
\bibitem [{\citenamefont {Berengut}\ and\ \citenamefont {Flambaum}(2010)}]{Berengut1}%
  \BibitemOpen
  \bibfield  {author} {\bibinfo {author} {\bibfnamefont {J.~C.}\ \bibnamefont {Berengut}}\ and\ \bibinfo {author} {\bibfnamefont {V.~V.}\ \bibnamefont {Flambaum}},\ }\href {https://doi.org/https://doi.org/10.1007/s10751-009-0146-y} {\bibfield  {journal} {\bibinfo  {journal} {Hyperfine Interactions}\ }\textbf {\bibinfo {volume} {196}},\ \bibinfo {pages} {269} (\bibinfo {year} {2010})}\BibitemShut {NoStop}%
\bibitem [{\citenamefont {Preskill}\ \emph {et~al.}(1983)\citenamefont {Preskill}, \citenamefont {Wise},\ and\ \citenamefont {Wilczek}}]{Preskill}%
  \BibitemOpen
  \bibfield  {author} {\bibinfo {author} {\bibfnamefont {J.}~\bibnamefont {Preskill}}, \bibinfo {author} {\bibfnamefont {M.~B.}\ \bibnamefont {Wise}},\ and\ \bibinfo {author} {\bibfnamefont {F.}~\bibnamefont {Wilczek}},\ }\href {https://doi.org/https://doi.org/10.1016/0370-2693(83)90637-8} {\bibfield  {journal} {\bibinfo  {journal} {Phys. Lett. B}\ }\textbf {\bibinfo {volume} {120}},\ \bibinfo {pages} {127} (\bibinfo {year} {1983})}\BibitemShut {NoStop}%
\bibitem [{\citenamefont {Abbott}\ and\ \citenamefont {Sikivie}(1983)}]{Abbott}%
  \BibitemOpen
  \bibfield  {author} {\bibinfo {author} {\bibfnamefont {L.}~\bibnamefont {Abbott}}\ and\ \bibinfo {author} {\bibfnamefont {P.}~\bibnamefont {Sikivie}},\ }\href {https://doi.org/https://doi.org/10.1016/0370-2693(83)90638-X} {\bibfield  {journal} {\bibinfo  {journal} {Phys. Lett. B}\ }\textbf {\bibinfo {volume} {120}},\ \bibinfo {pages} {133} (\bibinfo {year} {1983})}\BibitemShut {NoStop}%
\bibitem [{\citenamefont {Dine}\ and\ \citenamefont {Fischler}(1983)}]{Dine}%
  \BibitemOpen
  \bibfield  {author} {\bibinfo {author} {\bibfnamefont {M.}~\bibnamefont {Dine}}\ and\ \bibinfo {author} {\bibfnamefont {W.}~\bibnamefont {Fischler}},\ }\href {https://doi.org/https://doi.org/10.1016/0370-2693(83)90639-1} {\bibfield  {journal} {\bibinfo  {journal} {Phys. Lett. B}\ }\textbf {\bibinfo {volume} {120}},\ \bibinfo {pages} {137} (\bibinfo {year} {1983})}\BibitemShut {NoStop}%
\bibitem [{\citenamefont {Read}(2014)}]{DMdensity}%
  \BibitemOpen
  \bibfield  {author} {\bibinfo {author} {\bibfnamefont {J.~I.}\ \bibnamefont {Read}},\ }\href {https://doi.org/10.1088/0954-3899/41/6/063101} {\bibfield  {journal} {\bibinfo  {journal} {J. Phys. G}\ }\textbf {\bibinfo {volume} {41}},\ \bibinfo {pages} {063101} (\bibinfo {year} {2014})}\BibitemShut {NoStop}%
\bibitem [{\citenamefont {Arvanitaki}\ \emph {et~al.}(2015)\citenamefont {Arvanitaki}, \citenamefont {Huang},\ and\ \citenamefont {Van~Tilburg}}]{Arvanitaki}%
  \BibitemOpen
  \bibfield  {author} {\bibinfo {author} {\bibfnamefont {A.}~\bibnamefont {Arvanitaki}}, \bibinfo {author} {\bibfnamefont {J.}~\bibnamefont {Huang}},\ and\ \bibinfo {author} {\bibfnamefont {K.}~\bibnamefont {Van~Tilburg}},\ }\href {https://doi.org/10.1103/PhysRevD.91.015015} {\bibfield  {journal} {\bibinfo  {journal} {Phys. Rev. D}\ }\textbf {\bibinfo {volume} {91}},\ \bibinfo {pages} {015015} (\bibinfo {year} {2015})}\BibitemShut {NoStop}%
\bibitem [{\citenamefont {Stadnik}\ and\ \citenamefont {Flambaum}(2015{\natexlab{a}})}]{Stadnik}%
  \BibitemOpen
  \bibfield  {author} {\bibinfo {author} {\bibfnamefont {Y.~V.}\ \bibnamefont {Stadnik}}\ and\ \bibinfo {author} {\bibfnamefont {V.~V.}\ \bibnamefont {Flambaum}},\ }\href {https://doi.org/10.1103/PhysRevLett.115.201301} {\bibfield  {journal} {\bibinfo  {journal} {Phys. Rev. Lett.}\ }\textbf {\bibinfo {volume} {115}},\ \bibinfo {pages} {201301} (\bibinfo {year} {2015}{\natexlab{a}})}\BibitemShut {NoStop}%
\bibitem [{\citenamefont {Stadnik}\ and\ \citenamefont {Flambaum}(2015{\natexlab{b}})}]{Stadnik2015}%
  \BibitemOpen
  \bibfield  {author} {\bibinfo {author} {\bibfnamefont {Y.~V.}\ \bibnamefont {Stadnik}}\ and\ \bibinfo {author} {\bibfnamefont {V.~V.}\ \bibnamefont {Flambaum}},\ }\href {https://doi.org/10.1103/PhysRevLett.114.161301} {\bibfield  {journal} {\bibinfo  {journal} {Phys. Rev. Lett.}\ }\textbf {\bibinfo {volume} {114}},\ \bibinfo {pages} {161301} (\bibinfo {year} {2015}{\natexlab{b}})}\BibitemShut {NoStop}%
\bibitem [{\citenamefont {Kim}\ and\ \citenamefont {Perez}(2024)}]{KimPerez2024}%
  \BibitemOpen
  \bibfield  {author} {\bibinfo {author} {\bibfnamefont {H.}~\bibnamefont {Kim}}\ and\ \bibinfo {author} {\bibfnamefont {G.}~\bibnamefont {Perez}},\ }\href {https://doi.org/10.1103/PhysRevD.109.015005} {\bibfield  {journal} {\bibinfo  {journal} {Phys. Rev. D}\ }\textbf {\bibinfo {volume} {109}},\ \bibinfo {pages} {015005} (\bibinfo {year} {2024})}\BibitemShut {NoStop}%
\bibitem [{\citenamefont {Kim}\ \emph {et~al.}(2023)\citenamefont {Kim}, \citenamefont {Lenoci}, \citenamefont {Perez},\ and\ \citenamefont {Ratzinger}}]{kim2023probing}%
  \BibitemOpen
  \bibfield  {author} {\bibinfo {author} {\bibfnamefont {H.}~\bibnamefont {Kim}}, \bibinfo {author} {\bibfnamefont {A.}~\bibnamefont {Lenoci}}, \bibinfo {author} {\bibfnamefont {G.}~\bibnamefont {Perez}},\ and\ \bibinfo {author} {\bibfnamefont {W.}~\bibnamefont {Ratzinger}},\ }\href@noop {} {\bibinfo {title} {Probing an ultralight qcd axion with electromagnetic quadratic interaction}} (\bibinfo {year} {2023}),\ \Eprint {https://arxiv.org/abs/2307.14962} {arXiv:2307.14962 [hep-ph]} \BibitemShut {NoStop}%
\bibitem [{\citenamefont {Flambaum}\ and\ \citenamefont {Samsonov}(2023)}]{Samsonov}%
  \BibitemOpen
  \bibfield  {author} {\bibinfo {author} {\bibfnamefont {V.~V.}\ \bibnamefont {Flambaum}}\ and\ \bibinfo {author} {\bibfnamefont {I.~B.}\ \bibnamefont {Samsonov}},\ }\href {https://doi.org/10.1103/PhysRevD.108.075022} {\bibfield  {journal} {\bibinfo  {journal} {Phys. Rev. D}\ }\textbf {\bibinfo {volume} {108}},\ \bibinfo {pages} {075022} (\bibinfo {year} {2023})}\BibitemShut {NoStop}%
\bibitem [{\citenamefont {Dzuba}\ \emph {et~al.}(1999{\natexlab{a}})\citenamefont {Dzuba}, \citenamefont {Flambaum},\ and\ \citenamefont {Webb}}]{PRLWebb}%
  \BibitemOpen
  \bibfield  {author} {\bibinfo {author} {\bibfnamefont {V.~A.}\ \bibnamefont {Dzuba}}, \bibinfo {author} {\bibfnamefont {V.~V.}\ \bibnamefont {Flambaum}},\ and\ \bibinfo {author} {\bibfnamefont {J.~K.}\ \bibnamefont {Webb}},\ }\href {https://doi.org/10.1103/PhysRevLett.82.888} {\bibfield  {journal} {\bibinfo  {journal} {Phys. Rev. Lett.}\ }\textbf {\bibinfo {volume} {82}},\ \bibinfo {pages} {888} (\bibinfo {year} {1999}{\natexlab{a}})}\BibitemShut {NoStop}%
\bibitem [{\citenamefont {Dzuba}\ \emph {et~al.}(1999{\natexlab{b}})\citenamefont {Dzuba}, \citenamefont {Flambaum},\ and\ \citenamefont {Webb}}]{PRAWebb}%
  \BibitemOpen
  \bibfield  {author} {\bibinfo {author} {\bibfnamefont {V.~A.}\ \bibnamefont {Dzuba}}, \bibinfo {author} {\bibfnamefont {V.~V.}\ \bibnamefont {Flambaum}},\ and\ \bibinfo {author} {\bibfnamefont {J.~K.}\ \bibnamefont {Webb}},\ }\href {https://doi.org/10.1103/PhysRevA.59.230} {\bibfield  {journal} {\bibinfo  {journal} {Phys. Rev. A}\ }\textbf {\bibinfo {volume} {59}},\ \bibinfo {pages} {230} (\bibinfo {year} {1999}{\natexlab{b}})}\BibitemShut {NoStop}%
\bibitem [{\citenamefont {Flambaum}\ and\ \citenamefont {Dzuba}(2009)}]{CanJPh}%
  \BibitemOpen
  \bibfield  {author} {\bibinfo {author} {\bibfnamefont {V.~V.}\ \bibnamefont {Flambaum}}\ and\ \bibinfo {author} {\bibfnamefont {V.~A.}\ \bibnamefont {Dzuba}},\ }\href {https://doi.org/10.1139/p08-072} {\bibfield  {journal} {\bibinfo  {journal} {Can. J. Phys.}\ }\textbf {\bibinfo {volume} {87}},\ \bibinfo {pages} {25} (\bibinfo {year} {2009})}\BibitemShut {NoStop}%
\bibitem [{\citenamefont {Flambaum}\ and\ \citenamefont {Tedesco}(2006)}]{Tedesco2006}%
  \BibitemOpen
  \bibfield  {author} {\bibinfo {author} {\bibfnamefont {V.~V.}\ \bibnamefont {Flambaum}}\ and\ \bibinfo {author} {\bibfnamefont {A.~F.}\ \bibnamefont {Tedesco}},\ }\href {https://doi.org/10.1103/PhysRevC.73.055501} {\bibfield  {journal} {\bibinfo  {journal} {Phys. Rev. C}\ }\textbf {\bibinfo {volume} {73}},\ \bibinfo {pages} {055501} (\bibinfo {year} {2006})}\BibitemShut {NoStop}%
\bibitem [{\citenamefont {Pa\v{s}teka}\ \emph {et~al.}(2019)\citenamefont {Pa\v{s}teka}, \citenamefont {Hao}, \citenamefont {Borschevsky}, \citenamefont {Flambaum},\ and\ \citenamefont {Schwerdtfeger}}]{Borschevsky}%
  \BibitemOpen
  \bibfield  {author} {\bibinfo {author} {\bibfnamefont {L.~F.}\ \bibnamefont {Pa\v{s}teka}}, \bibinfo {author} {\bibfnamefont {Y.}~\bibnamefont {Hao}}, \bibinfo {author} {\bibfnamefont {A.}~\bibnamefont {Borschevsky}}, \bibinfo {author} {\bibfnamefont {V.~V.}\ \bibnamefont {Flambaum}},\ and\ \bibinfo {author} {\bibfnamefont {P.}~\bibnamefont {Schwerdtfeger}},\ }\href {https://doi.org/10.1103/PhysRevLett.122.160801} {\bibfield  {journal} {\bibinfo  {journal} {Phys. Rev. Lett.}\ }\textbf {\bibinfo {volume} {122}},\ \bibinfo {pages} {160801} (\bibinfo {year} {2019})}\BibitemShut {NoStop}%
\bibitem [{\citenamefont {Flambaum}\ and\ \citenamefont {Munro-Laylim}(2023)}]{csquarks}%
  \BibitemOpen
  \bibfield  {author} {\bibinfo {author} {\bibfnamefont {V.~V.}\ \bibnamefont {Flambaum}}\ and\ \bibinfo {author} {\bibfnamefont {P.}~\bibnamefont {Munro-Laylim}},\ }\href {https://doi.org/10.1103/PhysRevD.107.015004} {\bibfield  {journal} {\bibinfo  {journal} {Phys. Rev. D}\ }\textbf {\bibinfo {volume} {107}},\ \bibinfo {pages} {015004} (\bibinfo {year} {2023})}\BibitemShut {NoStop}%
\bibitem [{\citenamefont {Stadnik}\ and\ \citenamefont {Flambaum}(2016)}]{Stadnik2}%
  \BibitemOpen
  \bibfield  {author} {\bibinfo {author} {\bibfnamefont {Y.~V.}\ \bibnamefont {Stadnik}}\ and\ \bibinfo {author} {\bibfnamefont {V.~V.}\ \bibnamefont {Flambaum}},\ }\href {https://doi.org/10.1103/PhysRevA.94.022111} {\bibfield  {journal} {\bibinfo  {journal} {Phys. Rev. A}\ }\textbf {\bibinfo {volume} {94}},\ \bibinfo {pages} {022111} (\bibinfo {year} {2016})}\BibitemShut {NoStop}%
\bibitem [{\citenamefont {Van~Tilburg}\ \emph {et~al.}(2015)\citenamefont {Van~Tilburg}, \citenamefont {Leefer}, \citenamefont {Bougas},\ and\ \citenamefont {Budker}}]{DyCs}%
  \BibitemOpen
  \bibfield  {author} {\bibinfo {author} {\bibfnamefont {K.}~\bibnamefont {Van~Tilburg}}, \bibinfo {author} {\bibfnamefont {N.}~\bibnamefont {Leefer}}, \bibinfo {author} {\bibfnamefont {L.}~\bibnamefont {Bougas}},\ and\ \bibinfo {author} {\bibfnamefont {D.}~\bibnamefont {Budker}},\ }\href {https://doi.org/10.1103/PhysRevLett.115.011802} {\bibfield  {journal} {\bibinfo  {journal} {Phys. Rev. Lett.}\ }\textbf {\bibinfo {volume} {115}},\ \bibinfo {pages} {011802} (\bibinfo {year} {2015})}\BibitemShut {NoStop}%
\bibitem [{\citenamefont {Hees}\ \emph {et~al.}(2016)\citenamefont {Hees}, \citenamefont {Gu\'ena}, \citenamefont {Abgrall}, \citenamefont {Bize},\ and\ \citenamefont {Wolf}}]{Hees2016}%
  \BibitemOpen
  \bibfield  {author} {\bibinfo {author} {\bibfnamefont {A.}~\bibnamefont {Hees}}, \bibinfo {author} {\bibfnamefont {J.}~\bibnamefont {Gu\'ena}}, \bibinfo {author} {\bibfnamefont {M.}~\bibnamefont {Abgrall}}, \bibinfo {author} {\bibfnamefont {S.}~\bibnamefont {Bize}},\ and\ \bibinfo {author} {\bibfnamefont {P.}~\bibnamefont {Wolf}},\ }\href {https://doi.org/10.1103/PhysRevLett.117.061301} {\bibfield  {journal} {\bibinfo  {journal} {Phys. Rev. Lett.}\ }\textbf {\bibinfo {volume} {117}},\ \bibinfo {pages} {061301} (\bibinfo {year} {2016})}\BibitemShut {NoStop}%
\bibitem [{\citenamefont {Kobayashi}\ \emph {et~al.}(2022)\citenamefont {Kobayashi}, \citenamefont {Takamizawa}, \citenamefont {Akamatsu}, \citenamefont {Kawasaki}, \citenamefont {Nishiyama}, \citenamefont {Hosaka}, \citenamefont {Hisai}, \citenamefont {Wada}, \citenamefont {Inaba}, \citenamefont {Tanabe},\ and\ \citenamefont {Yasuda}}]{YbCs}%
  \BibitemOpen
  \bibfield  {author} {\bibinfo {author} {\bibfnamefont {T.}~\bibnamefont {Kobayashi}}, \bibinfo {author} {\bibfnamefont {A.}~\bibnamefont {Takamizawa}}, \bibinfo {author} {\bibfnamefont {D.}~\bibnamefont {Akamatsu}}, \bibinfo {author} {\bibfnamefont {A.}~\bibnamefont {Kawasaki}}, \bibinfo {author} {\bibfnamefont {A.}~\bibnamefont {Nishiyama}}, \bibinfo {author} {\bibfnamefont {K.}~\bibnamefont {Hosaka}}, \bibinfo {author} {\bibfnamefont {Y.}~\bibnamefont {Hisai}}, \bibinfo {author} {\bibfnamefont {M.}~\bibnamefont {Wada}}, \bibinfo {author} {\bibfnamefont {H.}~\bibnamefont {Inaba}}, \bibinfo {author} {\bibfnamefont {T.}~\bibnamefont {Tanabe}},\ and\ \bibinfo {author} {\bibfnamefont {M.}~\bibnamefont {Yasuda}},\ }\href {https://doi.org/10.1103/PhysRevLett.129.241301} {\bibfield  {journal} {\bibinfo  {journal} {Phys. Rev. Lett.}\ }\textbf {\bibinfo {volume} {129}},\ \bibinfo {pages} {241301} (\bibinfo {year} {2022})}\BibitemShut {NoStop}%
\bibitem [{\citenamefont {Kennedy}\ \emph {et~al.}(2020)\citenamefont {Kennedy}, \citenamefont {Oelker}, \citenamefont {Robinson}, \citenamefont {Bothwell}, \citenamefont {Kedar}, \citenamefont {Milner}, \citenamefont {Marti}, \citenamefont {Derevianko},\ and\ \citenamefont {Ye}}]{HSi}%
  \BibitemOpen
  \bibfield  {author} {\bibinfo {author} {\bibfnamefont {C.~J.}\ \bibnamefont {Kennedy}}, \bibinfo {author} {\bibfnamefont {E.}~\bibnamefont {Oelker}}, \bibinfo {author} {\bibfnamefont {J.~M.}\ \bibnamefont {Robinson}}, \bibinfo {author} {\bibfnamefont {T.}~\bibnamefont {Bothwell}}, \bibinfo {author} {\bibfnamefont {D.}~\bibnamefont {Kedar}}, \bibinfo {author} {\bibfnamefont {W.~R.}\ \bibnamefont {Milner}}, \bibinfo {author} {\bibfnamefont {G.~E.}\ \bibnamefont {Marti}}, \bibinfo {author} {\bibfnamefont {A.}~\bibnamefont {Derevianko}},\ and\ \bibinfo {author} {\bibfnamefont {J.}~\bibnamefont {Ye}},\ }\href {https://doi.org/10.1103/PhysRevLett.125.201302} {\bibfield  {journal} {\bibinfo  {journal} {Phys. Rev. Lett.}\ }\textbf {\bibinfo {volume} {125}},\ \bibinfo {pages} {201302} (\bibinfo {year} {2020})}\BibitemShut {NoStop}%
\bibitem [{\citenamefont {Tretiak}\ \emph {et~al.}(2022)\citenamefont {Tretiak}, \citenamefont {Zhang}, \citenamefont {Figueroa}, \citenamefont {Antypas}, \citenamefont {Brogna}, \citenamefont {Banerjee}, \citenamefont {Perez},\ and\ \citenamefont {Budker}}]{Tretiak}%
  \BibitemOpen
  \bibfield  {author} {\bibinfo {author} {\bibfnamefont {O.}~\bibnamefont {Tretiak}}, \bibinfo {author} {\bibfnamefont {X.}~\bibnamefont {Zhang}}, \bibinfo {author} {\bibfnamefont {N.~L.}\ \bibnamefont {Figueroa}}, \bibinfo {author} {\bibfnamefont {D.}~\bibnamefont {Antypas}}, \bibinfo {author} {\bibfnamefont {A.}~\bibnamefont {Brogna}}, \bibinfo {author} {\bibfnamefont {A.}~\bibnamefont {Banerjee}}, \bibinfo {author} {\bibfnamefont {G.}~\bibnamefont {Perez}},\ and\ \bibinfo {author} {\bibfnamefont {D.}~\bibnamefont {Budker}},\ }\href {https://doi.org/10.1103/PhysRevLett.129.031301} {\bibfield  {journal} {\bibinfo  {journal} {Phys. Rev. Lett.}\ }\textbf {\bibinfo {volume} {129}},\ \bibinfo {pages} {031301} (\bibinfo {year} {2022})}\BibitemShut {NoStop}%
\bibitem [{\citenamefont {Fischer}\ \emph {et~al.}(2004)\citenamefont {Fischer}, \citenamefont {Kolachevsky}, \citenamefont {Zimmermann}, \citenamefont {Holzwarth}, \citenamefont {Udem}, \citenamefont {H{\"a}nsch}, \citenamefont {Abgrall}, \citenamefont {Gr{\"u}nert}, \citenamefont {Maksimovic}, \citenamefont {Bize}, \citenamefont {Marion}, \citenamefont {Pereira Dos~Santos}, \citenamefont {Lemonde}, \citenamefont {Santarelli}, \citenamefont {Laurent}, \citenamefont {Clairon},\ and\ \citenamefont {Salomon}}]{Fischer2004}%
  \BibitemOpen
  \bibfield  {author} {\bibinfo {author} {\bibfnamefont {M.}~\bibnamefont {Fischer}}, \bibinfo {author} {\bibfnamefont {N.}~\bibnamefont {Kolachevsky}}, \bibinfo {author} {\bibfnamefont {M.}~\bibnamefont {Zimmermann}}, \bibinfo {author} {\bibfnamefont {R.}~\bibnamefont {Holzwarth}}, \bibinfo {author} {\bibfnamefont {T.}~\bibnamefont {Udem}}, \bibinfo {author} {\bibfnamefont {T.}~\bibnamefont {H{\"a}nsch}}, \bibinfo {author} {\bibfnamefont {M.}~\bibnamefont {Abgrall}}, \bibinfo {author} {\bibfnamefont {J.}~\bibnamefont {Gr{\"u}nert}}, \bibinfo {author} {\bibfnamefont {I.}~\bibnamefont {Maksimovic}}, \bibinfo {author} {\bibfnamefont {S.}~\bibnamefont {Bize}}, \bibinfo {author} {\bibfnamefont {H.}~\bibnamefont {Marion}}, \bibinfo {author} {\bibfnamefont {F.}~\bibnamefont {Pereira Dos~Santos}}, \bibinfo {author} {\bibfnamefont {P.}~\bibnamefont {Lemonde}}, \bibinfo {author} {\bibfnamefont {G.}~\bibnamefont {Santarelli}}, \bibinfo {author} {\bibfnamefont {P.}~\bibnamefont {Laurent}}, \bibinfo {author}
  {\bibfnamefont {A.}~\bibnamefont {Clairon}},\ and\ \bibinfo {author} {\bibfnamefont {C.}~\bibnamefont {Salomon}},\ }\bibinfo {title} {Precision spectroscopy of atomic hydrogen and variations of fundamental constants},\ in\ \href {https://doi.org/10.1007/978-3-540-40991-5_13} {\emph {\bibinfo {booktitle} {Astrophysics, Clocks and Fundamental Constants}}},\ \bibinfo {editor} {edited by\ \bibinfo {editor} {\bibfnamefont {S.~G.}\ \bibnamefont {Karshenboim}}\ and\ \bibinfo {editor} {\bibfnamefont {E.}~\bibnamefont {Peik}}}\ (\bibinfo  {publisher} {Springer Berlin Heidelberg},\ \bibinfo {address} {Berlin, Heidelberg},\ \bibinfo {year} {2004})\ pp.\ \bibinfo {pages} {209--227}\BibitemShut {NoStop}%
\bibitem [{\citenamefont {Rosenband}\ \emph {et~al.}(2008)\citenamefont {Rosenband}, \citenamefont {Hume}, \citenamefont {Schmidt}, \citenamefont {Chou}, \citenamefont {Brusch}, \citenamefont {Lorini}, \citenamefont {Oskay}, \citenamefont {Drullinger}, \citenamefont {Fortier}, \citenamefont {Stalnaker}, \citenamefont {Diddams}, \citenamefont {Swann}, \citenamefont {Newbury}, \citenamefont {Itano}, \citenamefont {Wineland},\ and\ \citenamefont {Bergquist}}]{AlHgdrift}%
  \BibitemOpen
  \bibfield  {author} {\bibinfo {author} {\bibfnamefont {T.}~\bibnamefont {Rosenband}}, \bibinfo {author} {\bibfnamefont {D.~B.}\ \bibnamefont {Hume}}, \bibinfo {author} {\bibfnamefont {P.~O.}\ \bibnamefont {Schmidt}}, \bibinfo {author} {\bibfnamefont {C.~W.}\ \bibnamefont {Chou}}, \bibinfo {author} {\bibfnamefont {A.}~\bibnamefont {Brusch}}, \bibinfo {author} {\bibfnamefont {L.}~\bibnamefont {Lorini}}, \bibinfo {author} {\bibfnamefont {W.~H.}\ \bibnamefont {Oskay}}, \bibinfo {author} {\bibfnamefont {R.~E.}\ \bibnamefont {Drullinger}}, \bibinfo {author} {\bibfnamefont {T.~M.}\ \bibnamefont {Fortier}}, \bibinfo {author} {\bibfnamefont {J.~E.}\ \bibnamefont {Stalnaker}}, \bibinfo {author} {\bibfnamefont {S.~A.}\ \bibnamefont {Diddams}}, \bibinfo {author} {\bibfnamefont {W.~C.}\ \bibnamefont {Swann}}, \bibinfo {author} {\bibfnamefont {N.~R.}\ \bibnamefont {Newbury}}, \bibinfo {author} {\bibfnamefont {W.~M.}\ \bibnamefont {Itano}}, \bibinfo {author} {\bibfnamefont {D.~J.}\ \bibnamefont {Wineland}},\ and\ \bibinfo
  {author} {\bibfnamefont {J.~C.}\ \bibnamefont {Bergquist}},\ }\href {https://doi.org/10.1126/science.1154622} {\bibfield  {journal} {\bibinfo  {journal} {Science}\ }\textbf {\bibinfo {volume} {319}},\ \bibinfo {pages} {1808} (\bibinfo {year} {2008})},\ \Eprint {https://arxiv.org/abs/https://www.science.org/doi/pdf/10.1126/science.1154622} {https://www.science.org/doi/pdf/10.1126/science.1154622} \BibitemShut {NoStop}%
\bibitem [{\citenamefont {Banerjee}\ \emph {et~al.}(2023)\citenamefont {Banerjee}, \citenamefont {Budker}, \citenamefont {Filzinger}, \citenamefont {Huntemann}, \citenamefont {Paz}, \citenamefont {Perez}, \citenamefont {Porsev},\ and\ \citenamefont {Safronova}}]{Banerjee2023}%
  \BibitemOpen
  \bibfield  {author} {\bibinfo {author} {\bibfnamefont {A.}~\bibnamefont {Banerjee}}, \bibinfo {author} {\bibfnamefont {D.}~\bibnamefont {Budker}}, \bibinfo {author} {\bibfnamefont {M.}~\bibnamefont {Filzinger}}, \bibinfo {author} {\bibfnamefont {N.}~\bibnamefont {Huntemann}}, \bibinfo {author} {\bibfnamefont {G.}~\bibnamefont {Paz}}, \bibinfo {author} {\bibfnamefont {G.}~\bibnamefont {Perez}}, \bibinfo {author} {\bibfnamefont {S.}~\bibnamefont {Porsev}},\ and\ \bibinfo {author} {\bibfnamefont {M.}~\bibnamefont {Safronova}},\ }\href@noop {} {\bibinfo {title} {Oscillating nuclear charge radii as sensors for ultralight dark matter}} (\bibinfo {year} {2023}),\ \Eprint {https://arxiv.org/abs/2301.10784} {arXiv:2301.10784 [hep-ph]} \BibitemShut {NoStop}%
\bibitem [{\citenamefont {Filzinger}\ \emph {et~al.}(2023)\citenamefont {Filzinger}, \citenamefont {D\"orscher}, \citenamefont {Lange}, \citenamefont {Klose}, \citenamefont {Steinel}, \citenamefont {Benkler}, \citenamefont {Peik}, \citenamefont {Lisdat},\ and\ \citenamefont {Huntemann}}]{Filzinger2023}%
  \BibitemOpen
  \bibfield  {author} {\bibinfo {author} {\bibfnamefont {M.}~\bibnamefont {Filzinger}}, \bibinfo {author} {\bibfnamefont {S.}~\bibnamefont {D\"orscher}}, \bibinfo {author} {\bibfnamefont {R.}~\bibnamefont {Lange}}, \bibinfo {author} {\bibfnamefont {J.}~\bibnamefont {Klose}}, \bibinfo {author} {\bibfnamefont {M.}~\bibnamefont {Steinel}}, \bibinfo {author} {\bibfnamefont {E.}~\bibnamefont {Benkler}}, \bibinfo {author} {\bibfnamefont {E.}~\bibnamefont {Peik}}, \bibinfo {author} {\bibfnamefont {C.}~\bibnamefont {Lisdat}},\ and\ \bibinfo {author} {\bibfnamefont {N.}~\bibnamefont {Huntemann}},\ }\href {https://doi.org/10.1103/PhysRevLett.130.253001} {\bibfield  {journal} {\bibinfo  {journal} {Phys. Rev. Lett.}\ }\textbf {\bibinfo {volume} {130}},\ \bibinfo {pages} {253001} (\bibinfo {year} {2023})}\BibitemShut {NoStop}%
\bibitem [{\citenamefont {Schlamminger}\ \emph {et~al.}(2008)\citenamefont {Schlamminger}, \citenamefont {Choi}, \citenamefont {Wagner}, \citenamefont {Gundlach},\ and\ \citenamefont {Adelberger}}]{Torsion1}%
  \BibitemOpen
  \bibfield  {author} {\bibinfo {author} {\bibfnamefont {S.}~\bibnamefont {Schlamminger}}, \bibinfo {author} {\bibfnamefont {K.-Y.}\ \bibnamefont {Choi}}, \bibinfo {author} {\bibfnamefont {T.~A.}\ \bibnamefont {Wagner}}, \bibinfo {author} {\bibfnamefont {J.~H.}\ \bibnamefont {Gundlach}},\ and\ \bibinfo {author} {\bibfnamefont {E.~G.}\ \bibnamefont {Adelberger}},\ }\href {https://doi.org/10.1103/PhysRevLett.100.041101} {\bibfield  {journal} {\bibinfo  {journal} {Phys. Rev. Lett.}\ }\textbf {\bibinfo {volume} {100}},\ \bibinfo {pages} {041101} (\bibinfo {year} {2008})}\BibitemShut {NoStop}%
\bibitem [{\citenamefont {Adelberger}\ \emph {et~al.}(2009)\citenamefont {Adelberger}, \citenamefont {Gundlach}, \citenamefont {Heckel}, \citenamefont {Hoedl},\ and\ \citenamefont {Schlamminger}}]{Torsion2}%
  \BibitemOpen
  \bibfield  {author} {\bibinfo {author} {\bibfnamefont {E.}~\bibnamefont {Adelberger}}, \bibinfo {author} {\bibfnamefont {J.}~\bibnamefont {Gundlach}}, \bibinfo {author} {\bibfnamefont {B.}~\bibnamefont {Heckel}}, \bibinfo {author} {\bibfnamefont {S.}~\bibnamefont {Hoedl}},\ and\ \bibinfo {author} {\bibfnamefont {S.}~\bibnamefont {Schlamminger}},\ }\href {https://doi.org/https://doi.org/10.1016/j.ppnp.2008.08.002} {\bibfield  {journal} {\bibinfo  {journal} {Progress in Particle and Nuclear Physics}\ }\textbf {\bibinfo {volume} {62}},\ \bibinfo {pages} {102} (\bibinfo {year} {2009})}\BibitemShut {NoStop}%
\bibitem [{\citenamefont {Williams}\ \emph {et~al.}(2004)\citenamefont {Williams}, \citenamefont {Turyshev},\ and\ \citenamefont {Boggs}}]{LLRTests}%
  \BibitemOpen
  \bibfield  {author} {\bibinfo {author} {\bibfnamefont {J.~G.}\ \bibnamefont {Williams}}, \bibinfo {author} {\bibfnamefont {S.~G.}\ \bibnamefont {Turyshev}},\ and\ \bibinfo {author} {\bibfnamefont {D.~H.}\ \bibnamefont {Boggs}},\ }\href {https://doi.org/10.1103/PhysRevLett.93.261101} {\bibfield  {journal} {\bibinfo  {journal} {Phys. Rev. Lett.}\ }\textbf {\bibinfo {volume} {93}},\ \bibinfo {pages} {261101} (\bibinfo {year} {2004})}\BibitemShut {NoStop}%
\bibitem [{\citenamefont {Zhou}\ \emph {et~al.}(2015)\citenamefont {Zhou}, \citenamefont {Long}, \citenamefont {Tang}, \citenamefont {Chen}, \citenamefont {Gao}, \citenamefont {Peng}, \citenamefont {Duan}, \citenamefont {Zhong}, \citenamefont {Xiong}, \citenamefont {Wang}, \citenamefont {Zhang},\ and\ \citenamefont {Zhan}}]{Zhou2015}%
  \BibitemOpen
  \bibfield  {author} {\bibinfo {author} {\bibfnamefont {L.}~\bibnamefont {Zhou}}, \bibinfo {author} {\bibfnamefont {S.}~\bibnamefont {Long}}, \bibinfo {author} {\bibfnamefont {B.}~\bibnamefont {Tang}}, \bibinfo {author} {\bibfnamefont {X.}~\bibnamefont {Chen}}, \bibinfo {author} {\bibfnamefont {F.}~\bibnamefont {Gao}}, \bibinfo {author} {\bibfnamefont {W.}~\bibnamefont {Peng}}, \bibinfo {author} {\bibfnamefont {W.}~\bibnamefont {Duan}}, \bibinfo {author} {\bibfnamefont {J.}~\bibnamefont {Zhong}}, \bibinfo {author} {\bibfnamefont {Z.}~\bibnamefont {Xiong}}, \bibinfo {author} {\bibfnamefont {J.}~\bibnamefont {Wang}}, \bibinfo {author} {\bibfnamefont {Y.}~\bibnamefont {Zhang}},\ and\ \bibinfo {author} {\bibfnamefont {M.}~\bibnamefont {Zhan}},\ }\href {https://doi.org/10.1103/PhysRevLett.115.013004} {\bibfield  {journal} {\bibinfo  {journal} {Phys. Rev. Lett.}\ }\textbf {\bibinfo {volume} {115}},\ \bibinfo {pages} {013004} (\bibinfo {year} {2015})}\BibitemShut {NoStop}%
\bibitem [{\citenamefont {Berg\'e}\ \emph {et~al.}(2018)\citenamefont {Berg\'e}, \citenamefont {Brax}, \citenamefont {M\'etris}, \citenamefont {Pernot-Borr\`as}, \citenamefont {Touboul},\ and\ \citenamefont {Uzan}}]{MICROSCOPE1}%
  \BibitemOpen
  \bibfield  {author} {\bibinfo {author} {\bibfnamefont {J.}~\bibnamefont {Berg\'e}}, \bibinfo {author} {\bibfnamefont {P.}~\bibnamefont {Brax}}, \bibinfo {author} {\bibfnamefont {G.}~\bibnamefont {M\'etris}}, \bibinfo {author} {\bibfnamefont {M.}~\bibnamefont {Pernot-Borr\`as}}, \bibinfo {author} {\bibfnamefont {P.}~\bibnamefont {Touboul}},\ and\ \bibinfo {author} {\bibfnamefont {J.-P.}\ \bibnamefont {Uzan}},\ }\href {https://doi.org/10.1103/PhysRevLett.120.141101} {\bibfield  {journal} {\bibinfo  {journal} {Phys. Rev. Lett.}\ }\textbf {\bibinfo {volume} {120}},\ \bibinfo {pages} {141101} (\bibinfo {year} {2018})}\BibitemShut {NoStop}%
\bibitem [{\citenamefont {Touboul}\ \emph {et~al.}(2017)\citenamefont {Touboul}, \citenamefont {M\'etris}, \citenamefont {Rodrigues}, \citenamefont {Andr\'e}, \citenamefont {Baghi}, \citenamefont {Berg\'e}, \citenamefont {Boulanger}, \citenamefont {Bremer}, \citenamefont {Carle}, \citenamefont {Chhun}, \citenamefont {Christophe}, \citenamefont {Cipolla}, \citenamefont {Damour}, \citenamefont {Danto}, \citenamefont {Dittus}, \citenamefont {Fayet}, \citenamefont {Foulon}, \citenamefont {Gageant}, \citenamefont {Guidotti}, \citenamefont {Hagedorn}, \citenamefont {Hardy}, \citenamefont {Huynh}, \citenamefont {Inchauspe}, \citenamefont {Kayser}, \citenamefont {Lala}, \citenamefont {L\"ammerzahl}, \citenamefont {Lebat}, \citenamefont {Leseur}, \citenamefont {Liorzou}, \citenamefont {List}, \citenamefont {L\"offler}, \citenamefont {Panet}, \citenamefont {Pouilloux}, \citenamefont {Prieur}, \citenamefont {Rebray}, \citenamefont {Reynaud}, \citenamefont {Rievers}, \citenamefont {Robert}, \citenamefont {Selig},
  \citenamefont {Serron}, \citenamefont {Sumner}, \citenamefont {Tanguy},\ and\ \citenamefont {Visser}}]{MICROSCOPE2}%
  \BibitemOpen
  \bibfield  {author} {\bibinfo {author} {\bibfnamefont {P.}~\bibnamefont {Touboul}}, \bibinfo {author} {\bibfnamefont {G.}~\bibnamefont {M\'etris}}, \bibinfo {author} {\bibfnamefont {M.}~\bibnamefont {Rodrigues}}, \bibinfo {author} {\bibfnamefont {Y.}~\bibnamefont {Andr\'e}}, \bibinfo {author} {\bibfnamefont {Q.}~\bibnamefont {Baghi}}, \bibinfo {author} {\bibfnamefont {J.}~\bibnamefont {Berg\'e}}, \bibinfo {author} {\bibfnamefont {D.}~\bibnamefont {Boulanger}}, \bibinfo {author} {\bibfnamefont {S.}~\bibnamefont {Bremer}}, \bibinfo {author} {\bibfnamefont {P.}~\bibnamefont {Carle}}, \bibinfo {author} {\bibfnamefont {R.}~\bibnamefont {Chhun}}, \bibinfo {author} {\bibfnamefont {B.}~\bibnamefont {Christophe}}, \bibinfo {author} {\bibfnamefont {V.}~\bibnamefont {Cipolla}}, \bibinfo {author} {\bibfnamefont {T.}~\bibnamefont {Damour}}, \bibinfo {author} {\bibfnamefont {P.}~\bibnamefont {Danto}}, \bibinfo {author} {\bibfnamefont {H.}~\bibnamefont {Dittus}}, \bibinfo {author} {\bibfnamefont {P.}~\bibnamefont
  {Fayet}}, \bibinfo {author} {\bibfnamefont {B.}~\bibnamefont {Foulon}}, \bibinfo {author} {\bibfnamefont {C.}~\bibnamefont {Gageant}}, \bibinfo {author} {\bibfnamefont {P.-Y.}\ \bibnamefont {Guidotti}}, \bibinfo {author} {\bibfnamefont {D.}~\bibnamefont {Hagedorn}}, \bibinfo {author} {\bibfnamefont {E.}~\bibnamefont {Hardy}}, \bibinfo {author} {\bibfnamefont {P.-A.}\ \bibnamefont {Huynh}}, \bibinfo {author} {\bibfnamefont {H.}~\bibnamefont {Inchauspe}}, \bibinfo {author} {\bibfnamefont {P.}~\bibnamefont {Kayser}}, \bibinfo {author} {\bibfnamefont {S.}~\bibnamefont {Lala}}, \bibinfo {author} {\bibfnamefont {C.}~\bibnamefont {L\"ammerzahl}}, \bibinfo {author} {\bibfnamefont {V.}~\bibnamefont {Lebat}}, \bibinfo {author} {\bibfnamefont {P.}~\bibnamefont {Leseur}}, \bibinfo {author} {\bibfnamefont {F.~m.~c.}\ \bibnamefont {Liorzou}}, \bibinfo {author} {\bibfnamefont {M.}~\bibnamefont {List}}, \bibinfo {author} {\bibfnamefont {F.}~\bibnamefont {L\"offler}}, \bibinfo {author} {\bibfnamefont {I.}~\bibnamefont
  {Panet}}, \bibinfo {author} {\bibfnamefont {B.}~\bibnamefont {Pouilloux}}, \bibinfo {author} {\bibfnamefont {P.}~\bibnamefont {Prieur}}, \bibinfo {author} {\bibfnamefont {A.}~\bibnamefont {Rebray}}, \bibinfo {author} {\bibfnamefont {S.}~\bibnamefont {Reynaud}}, \bibinfo {author} {\bibfnamefont {B.}~\bibnamefont {Rievers}}, \bibinfo {author} {\bibfnamefont {A.}~\bibnamefont {Robert}}, \bibinfo {author} {\bibfnamefont {H.}~\bibnamefont {Selig}}, \bibinfo {author} {\bibfnamefont {L.}~\bibnamefont {Serron}}, \bibinfo {author} {\bibfnamefont {T.}~\bibnamefont {Sumner}}, \bibinfo {author} {\bibfnamefont {N.}~\bibnamefont {Tanguy}},\ and\ \bibinfo {author} {\bibfnamefont {P.}~\bibnamefont {Visser}},\ }\href {https://doi.org/10.1103/PhysRevLett.119.231101} {\bibfield  {journal} {\bibinfo  {journal} {Phys. Rev. Lett.}\ }\textbf {\bibinfo {volume} {119}},\ \bibinfo {pages} {231101} (\bibinfo {year} {2017})}\BibitemShut {NoStop}%
\bibitem [{\citenamefont {Leefer}\ \emph {et~al.}(2016)\citenamefont {Leefer}, \citenamefont {Gerhardus}, \citenamefont {Budker}, \citenamefont {Flambaum},\ and\ \citenamefont {Stadnik}}]{Leefer2016}%
  \BibitemOpen
  \bibfield  {author} {\bibinfo {author} {\bibfnamefont {N.}~\bibnamefont {Leefer}}, \bibinfo {author} {\bibfnamefont {A.}~\bibnamefont {Gerhardus}}, \bibinfo {author} {\bibfnamefont {D.}~\bibnamefont {Budker}}, \bibinfo {author} {\bibfnamefont {V.~V.}\ \bibnamefont {Flambaum}},\ and\ \bibinfo {author} {\bibfnamefont {Y.~V.}\ \bibnamefont {Stadnik}},\ }\href {https://doi.org/10.1103/PhysRevLett.117.271601} {\bibfield  {journal} {\bibinfo  {journal} {Phys. Rev. Lett.}\ }\textbf {\bibinfo {volume} {117}},\ \bibinfo {pages} {271601} (\bibinfo {year} {2016})}\BibitemShut {NoStop}%
\bibitem [{\citenamefont {Brzeminski}\ \emph {et~al.}(2022)\citenamefont {Brzeminski}, \citenamefont {Chacko}, \citenamefont {Dev}, \citenamefont {Flood},\ and\ \citenamefont {Hook}}]{Brzeminski2022}%
  \BibitemOpen
  \bibfield  {author} {\bibinfo {author} {\bibfnamefont {D.}~\bibnamefont {Brzeminski}}, \bibinfo {author} {\bibfnamefont {Z.}~\bibnamefont {Chacko}}, \bibinfo {author} {\bibfnamefont {A.}~\bibnamefont {Dev}}, \bibinfo {author} {\bibfnamefont {I.}~\bibnamefont {Flood}},\ and\ \bibinfo {author} {\bibfnamefont {A.}~\bibnamefont {Hook}},\ }\href {https://doi.org/10.1103/PhysRevD.106.095031} {\bibfield  {journal} {\bibinfo  {journal} {Phys. Rev. D}\ }\textbf {\bibinfo {volume} {106}},\ \bibinfo {pages} {095031} (\bibinfo {year} {2022})}\BibitemShut {NoStop}%
\bibitem [{\citenamefont {Dinh}\ \emph {et~al.}(2009)\citenamefont {Dinh}, \citenamefont {Dunning}, \citenamefont {Dzuba},\ and\ \citenamefont {Flambaum}}]{Dinh2009}%
  \BibitemOpen
  \bibfield  {author} {\bibinfo {author} {\bibfnamefont {T.~H.}\ \bibnamefont {Dinh}}, \bibinfo {author} {\bibfnamefont {A.}~\bibnamefont {Dunning}}, \bibinfo {author} {\bibfnamefont {V.~A.}\ \bibnamefont {Dzuba}},\ and\ \bibinfo {author} {\bibfnamefont {V.~V.}\ \bibnamefont {Flambaum}},\ }\href {https://doi.org/10.1103/physreva.79.054102} {\bibfield  {journal} {\bibinfo  {journal} {Phys. Rev. A}\ }\textbf {\bibinfo {volume} {79}},\ \bibinfo {pages} {054102} (\bibinfo {year} {2009})}\BibitemShut {NoStop}%
\bibitem [{\citenamefont {Flambaum}\ and\ \citenamefont {Mansour}(2023)}]{FlambaumMansour2023}%
  \BibitemOpen
  \bibfield  {author} {\bibinfo {author} {\bibfnamefont {V.~V.}\ \bibnamefont {Flambaum}}\ and\ \bibinfo {author} {\bibfnamefont {A.~J.}\ \bibnamefont {Mansour}},\ }\href {https://doi.org/10.1103/PhysRevLett.131.113004} {\bibfield  {journal} {\bibinfo  {journal} {Phys. Rev. Lett.}\ }\textbf {\bibinfo {volume} {131}},\ \bibinfo {pages} {113004} (\bibinfo {year} {2023})}\BibitemShut {NoStop}%
\bibitem [{\citenamefont {Colladay}\ and\ \citenamefont {Kosteleck\'y}(1998)}]{Kostelecky}%
  \BibitemOpen
  \bibfield  {author} {\bibinfo {author} {\bibfnamefont {D.}~\bibnamefont {Colladay}}\ and\ \bibinfo {author} {\bibfnamefont {V.~A.}\ \bibnamefont {Kosteleck\'y}},\ }\href {https://doi.org/10.1103/PhysRevD.58.116002} {\bibfield  {journal} {\bibinfo  {journal} {Phys. Rev. D}\ }\textbf {\bibinfo {volume} {58}},\ \bibinfo {pages} {116002} (\bibinfo {year} {1998})}\BibitemShut {NoStop}%
\bibitem [{\citenamefont {Lange}\ \emph {et~al.}(2021)\citenamefont {Lange}, \citenamefont {Huntemann}, \citenamefont {Rahm}, \citenamefont {Sanner}, \citenamefont {Shao}, \citenamefont {Lipphardt}, \citenamefont {Tamm}, \citenamefont {Weyers},\ and\ \citenamefont {Peik}}]{Lange2021}%
  \BibitemOpen
  \bibfield  {author} {\bibinfo {author} {\bibfnamefont {R.}~\bibnamefont {Lange}}, \bibinfo {author} {\bibfnamefont {N.}~\bibnamefont {Huntemann}}, \bibinfo {author} {\bibfnamefont {J.~M.}\ \bibnamefont {Rahm}}, \bibinfo {author} {\bibfnamefont {C.}~\bibnamefont {Sanner}}, \bibinfo {author} {\bibfnamefont {H.}~\bibnamefont {Shao}}, \bibinfo {author} {\bibfnamefont {B.}~\bibnamefont {Lipphardt}}, \bibinfo {author} {\bibfnamefont {C.}~\bibnamefont {Tamm}}, \bibinfo {author} {\bibfnamefont {S.}~\bibnamefont {Weyers}},\ and\ \bibinfo {author} {\bibfnamefont {E.}~\bibnamefont {Peik}},\ }\href {https://doi.org/10.1103/PhysRevLett.126.011102} {\bibfield  {journal} {\bibinfo  {journal} {Phys. Rev. Lett.}\ }\textbf {\bibinfo {volume} {126}},\ \bibinfo {pages} {011102} (\bibinfo {year} {2021})}\BibitemShut {NoStop}%
\bibitem [{\citenamefont {Flambaum}\ and\ \citenamefont {Shuryak}(2003)}]{FlambaumShuryak}%
  \BibitemOpen
  \bibfield  {author} {\bibinfo {author} {\bibfnamefont {V.~V.}\ \bibnamefont {Flambaum}}\ and\ \bibinfo {author} {\bibfnamefont {E.~V.}\ \bibnamefont {Shuryak}},\ }\href {https://doi.org/10.1103/PhysRevD.67.083507} {\bibfield  {journal} {\bibinfo  {journal} {Phys. Rev. D}\ }\textbf {\bibinfo {volume} {67}},\ \bibinfo {pages} {083507} (\bibinfo {year} {2003})}\BibitemShut {NoStop}%
\bibitem [{\citenamefont {Flambaum}\ and\ \citenamefont {Wiringa}(2009)}]{Wiringa2009}%
  \BibitemOpen
  \bibfield  {author} {\bibinfo {author} {\bibfnamefont {V.~V.}\ \bibnamefont {Flambaum}}\ and\ \bibinfo {author} {\bibfnamefont {R.~B.}\ \bibnamefont {Wiringa}},\ }\href {https://doi.org/10.1103/PhysRevC.79.034302} {\bibfield  {journal} {\bibinfo  {journal} {Phys. Rev. C}\ }\textbf {\bibinfo {volume} {79}},\ \bibinfo {pages} {034302} (\bibinfo {year} {2009})}\BibitemShut {NoStop}%
\bibitem [{\citenamefont {Krane}(2008)}]{Krane}%
  \BibitemOpen
  \bibfield  {author} {\bibinfo {author} {\bibfnamefont {K.}~\bibnamefont {Krane}},\ }\href {https://books.google.com.au/books?id=mkZjC1TLXh8C} {\emph {\bibinfo {title} {Introductory Nuclear Physics}}}\ (\bibinfo  {publisher} {Wiley India},\ \bibinfo {year} {2008})\BibitemShut {NoStop}%
\bibitem [{\citenamefont {King}(2013)}]{king2013isotope}%
  \BibitemOpen
  \bibfield  {author} {\bibinfo {author} {\bibfnamefont {W.}~\bibnamefont {King}},\ }\href {https://books.google.de/books?id=eEgGCAAAQBAJ} {\emph {\bibinfo {title} {Isotope Shifts in Atomic Spectra}}},\ Physics of Atoms and Molecules\ (\bibinfo  {publisher} {Springer US},\ \bibinfo {year} {2013})\BibitemShut {NoStop}%
\bibitem [{\citenamefont {Flambaum}\ \emph {et~al.}(2004)\citenamefont {Flambaum}, \citenamefont {Leinweber}, \citenamefont {Thomas},\ and\ \citenamefont {Young}}]{Flambaum2004}%
  \BibitemOpen
  \bibfield  {author} {\bibinfo {author} {\bibfnamefont {V.~V.}\ \bibnamefont {Flambaum}}, \bibinfo {author} {\bibfnamefont {D.~B.}\ \bibnamefont {Leinweber}}, \bibinfo {author} {\bibfnamefont {A.~W.}\ \bibnamefont {Thomas}},\ and\ \bibinfo {author} {\bibfnamefont {R.~D.}\ \bibnamefont {Young}},\ }\href {https://doi.org/10.1103/PhysRevD.69.115006} {\bibfield  {journal} {\bibinfo  {journal} {Phys. Rev. D}\ }\textbf {\bibinfo {volume} {69}},\ \bibinfo {pages} {115006} (\bibinfo {year} {2004})}\BibitemShut {NoStop}%
\bibitem [{\citenamefont {Ubaldi}(2010)}]{Ubaldi2010}%
  \BibitemOpen
  \bibfield  {author} {\bibinfo {author} {\bibfnamefont {L.}~\bibnamefont {Ubaldi}},\ }\href {https://doi.org/10.1103/PhysRevD.81.025011} {\bibfield  {journal} {\bibinfo  {journal} {Phys. Rev. D}\ }\textbf {\bibinfo {volume} {81}},\ \bibinfo {pages} {025011} (\bibinfo {year} {2010})}\BibitemShut {NoStop}%
\bibitem [{\citenamefont {Rosenband}\ \emph {et~al.}(2007)\citenamefont {Rosenband}, \citenamefont {Schmidt}, \citenamefont {Hume}, \citenamefont {Itano}, \citenamefont {Fortier}, \citenamefont {Stalnaker}, \citenamefont {Kim}, \citenamefont {Diddams}, \citenamefont {Koelemeij}, \citenamefont {Bergquist},\ and\ \citenamefont {Wineland}}]{Wineland2007}%
  \BibitemOpen
  \bibfield  {author} {\bibinfo {author} {\bibfnamefont {T.}~\bibnamefont {Rosenband}}, \bibinfo {author} {\bibfnamefont {P.~O.}\ \bibnamefont {Schmidt}}, \bibinfo {author} {\bibfnamefont {D.~B.}\ \bibnamefont {Hume}}, \bibinfo {author} {\bibfnamefont {W.~M.}\ \bibnamefont {Itano}}, \bibinfo {author} {\bibfnamefont {T.~M.}\ \bibnamefont {Fortier}}, \bibinfo {author} {\bibfnamefont {J.~E.}\ \bibnamefont {Stalnaker}}, \bibinfo {author} {\bibfnamefont {K.}~\bibnamefont {Kim}}, \bibinfo {author} {\bibfnamefont {S.~A.}\ \bibnamefont {Diddams}}, \bibinfo {author} {\bibfnamefont {J.~C.~J.}\ \bibnamefont {Koelemeij}}, \bibinfo {author} {\bibfnamefont {J.~C.}\ \bibnamefont {Bergquist}},\ and\ \bibinfo {author} {\bibfnamefont {D.~J.}\ \bibnamefont {Wineland}},\ }\href {https://doi.org/10.1103/PhysRevLett.98.220801} {\bibfield  {journal} {\bibinfo  {journal} {Phys. Rev. Lett.}\ }\textbf {\bibinfo {volume} {98}},\ \bibinfo {pages} {220801} (\bibinfo {year} {2007})}\BibitemShut {NoStop}%
\bibitem [{\citenamefont {Tang}\ \emph {et~al.}(2021)\citenamefont {Tang}, \citenamefont {Zhang}, \citenamefont {Shen},\ and\ \citenamefont {Zou}}]{Tang2021}%
  \BibitemOpen
  \bibfield  {author} {\bibinfo {author} {\bibfnamefont {X.-K.}\ \bibnamefont {Tang}}, \bibinfo {author} {\bibfnamefont {X.}~\bibnamefont {Zhang}}, \bibinfo {author} {\bibfnamefont {Y.}~\bibnamefont {Shen}},\ and\ \bibinfo {author} {\bibfnamefont {H.-X.}\ \bibnamefont {Zou}},\ }\href {https://doi.org/10.1088/1674-1056/ac0130} {\bibfield  {journal} {\bibinfo  {journal} {Chinese Physics B}\ }\textbf {\bibinfo {volume} {30}},\ \bibinfo {pages} {123204} (\bibinfo {year} {2021})}\BibitemShut {NoStop}%
\bibitem [{\citenamefont {Dzuba}\ \emph {et~al.}(2017)\citenamefont {Dzuba}, \citenamefont {Berengut}, \citenamefont {Harabati},\ and\ \citenamefont {Flambaum}}]{CIPT}%
  \BibitemOpen
  \bibfield  {author} {\bibinfo {author} {\bibfnamefont {V.~A.}\ \bibnamefont {Dzuba}}, \bibinfo {author} {\bibfnamefont {J.~C.}\ \bibnamefont {Berengut}}, \bibinfo {author} {\bibfnamefont {C.}~\bibnamefont {Harabati}},\ and\ \bibinfo {author} {\bibfnamefont {V.~V.}\ \bibnamefont {Flambaum}},\ }\href {https://doi.org/10.1103/PhysRevA.95.012503} {\bibfield  {journal} {\bibinfo  {journal} {Phys. Rev. A}\ }\textbf {\bibinfo {volume} {95}},\ \bibinfo {pages} {012503} (\bibinfo {year} {2017})}\BibitemShut {NoStop}%
\bibitem [{\citenamefont {Allehabi}\ \emph {et~al.}(2020)\citenamefont {Allehabi}, \citenamefont {Dzuba}, \citenamefont {Flambaum}, \citenamefont {Afanasjev},\ and\ \citenamefont {Agbemava}}]{NoIS}%
  \BibitemOpen
  \bibfield  {author} {\bibinfo {author} {\bibfnamefont {S.~O.}\ \bibnamefont {Allehabi}}, \bibinfo {author} {\bibfnamefont {V.~A.}\ \bibnamefont {Dzuba}}, \bibinfo {author} {\bibfnamefont {V.~V.}\ \bibnamefont {Flambaum}}, \bibinfo {author} {\bibfnamefont {A.~V.}\ \bibnamefont {Afanasjev}},\ and\ \bibinfo {author} {\bibfnamefont {S.~E.}\ \bibnamefont {Agbemava}},\ }\href {https://doi.org/10.1103/PhysRevC.102.024326} {\bibfield  {journal} {\bibinfo  {journal} {Phys. Rev. C}\ }\textbf {\bibinfo {volume} {102}},\ \bibinfo {pages} {024326} (\bibinfo {year} {2020})}\BibitemShut {NoStop}%
\bibitem [{\citenamefont {Xiang}\ \emph {et~al.}(2019)\citenamefont {Xiang}, \citenamefont {Ben-Quan}, \citenamefont {Ji-Guang},\ and\ \citenamefont {Hong-Xin}}]{Zhang2019}%
  \BibitemOpen
  \bibfield  {author} {\bibinfo {author} {\bibfnamefont {Z.}~\bibnamefont {Xiang}}, \bibinfo {author} {\bibfnamefont {L.}~\bibnamefont {Ben-Quan}}, \bibinfo {author} {\bibfnamefont {L.}~\bibnamefont {Ji-Guang}},\ and\ \bibinfo {author} {\bibfnamefont {Z.}~\bibnamefont {Hong-Xin}},\ }\bibfield  {journal} {\bibinfo  {journal} {Acta. Phys. Sin.}\ }\textbf {\bibinfo {volume} {68}},\ \href {https://doi.org/10.7498/aps.68.20182136} {10.7498/aps.68.20182136} (\bibinfo {year} {2019})\BibitemShut {NoStop}%
\bibitem [{\citenamefont {Angeli}\ and\ \citenamefont {Marinova}(2013)}]{Angeli2013}%
  \BibitemOpen
  \bibfield  {author} {\bibinfo {author} {\bibfnamefont {I.}~\bibnamefont {Angeli}}\ and\ \bibinfo {author} {\bibfnamefont {K.}~\bibnamefont {Marinova}},\ }\href {https://doi.org/https://doi.org/10.1016/j.adt.2011.12.006} {\bibfield  {journal} {\bibinfo  {journal} {Atomic Data and Nuclear Data Tables}\ }\textbf {\bibinfo {volume} {99}},\ \bibinfo {pages} {69} (\bibinfo {year} {2013})}\BibitemShut {NoStop}%
\bibitem [{\citenamefont {Flambaum}\ and\ \citenamefont {Wiringa}(2007)}]{Wiringa2007}%
  \BibitemOpen
  \bibfield  {author} {\bibinfo {author} {\bibfnamefont {V.~V.}\ \bibnamefont {Flambaum}}\ and\ \bibinfo {author} {\bibfnamefont {R.~B.}\ \bibnamefont {Wiringa}},\ }\href {https://doi.org/10.1103/PhysRevC.76.054002} {\bibfield  {journal} {\bibinfo  {journal} {Phys. Rev. C}\ }\textbf {\bibinfo {volume} {76}},\ \bibinfo {pages} {054002} (\bibinfo {year} {2007})}\BibitemShut {NoStop}%
\bibitem [{\citenamefont {{Clo{\"e}t}}\ \emph {et~al.}(2008)\citenamefont {{Clo{\"e}t}}, \citenamefont {{Eichmann}}, \citenamefont {{Flambaum}}, \citenamefont {{Roberts}}, \citenamefont {{Bhagwat}},\ and\ \citenamefont {{H{\"o}ll}}}]{Roberts2008}%
  \BibitemOpen
  \bibfield  {author} {\bibinfo {author} {\bibfnamefont {I.~C.}\ \bibnamefont {{Clo{\"e}t}}}, \bibinfo {author} {\bibfnamefont {G.}~\bibnamefont {{Eichmann}}}, \bibinfo {author} {\bibfnamefont {V.~V.}\ \bibnamefont {{Flambaum}}}, \bibinfo {author} {\bibfnamefont {C.~D.}\ \bibnamefont {{Roberts}}}, \bibinfo {author} {\bibfnamefont {M.~S.}\ \bibnamefont {{Bhagwat}}},\ and\ \bibinfo {author} {\bibfnamefont {A.}~\bibnamefont {{H{\"o}ll}}},\ }\href {https://doi.org/10.1007/s00601-008-0240-8} {\bibfield  {journal} {\bibinfo  {journal} {Few-Body Systems}\ }\textbf {\bibinfo {volume} {42}},\ \bibinfo {pages} {91} (\bibinfo {year} {2008})},\ \Eprint {https://arxiv.org/abs/0804.3118} {arXiv:0804.3118 [nucl-th]} \BibitemShut {NoStop}%
\bibitem [{\citenamefont {Flambaum}\ \emph {et~al.}(2006)\citenamefont {Flambaum}, \citenamefont {H\"{o}ll}, \citenamefont {Jaikumar}, \citenamefont {Roberts},\ and\ \citenamefont {Wright}}]{Flambaum2006}%
  \BibitemOpen
  \bibfield  {author} {\bibinfo {author} {\bibfnamefont {V.~V.}\ \bibnamefont {Flambaum}}, \bibinfo {author} {\bibfnamefont {A.}~\bibnamefont {H\"{o}ll}}, \bibinfo {author} {\bibfnamefont {P.}~\bibnamefont {Jaikumar}}, \bibinfo {author} {\bibfnamefont {C.~D.}\ \bibnamefont {Roberts}},\ and\ \bibinfo {author} {\bibfnamefont {S.~V.}\ \bibnamefont {Wright}},\ }\href {https://doi.org/10.1007/s00601-005-0123-1} {\bibfield  {journal} {\bibinfo  {journal} {Few-Body Systems}\ }\textbf {\bibinfo {volume} {38}},\ \bibinfo {pages} {31} (\bibinfo {year} {2006})}\BibitemShut {NoStop}%
\bibitem [{\citenamefont {Höll}\ \emph {et~al.}(2006)\citenamefont {Höll}, \citenamefont {Maris}, \citenamefont {Roberts},\ and\ \citenamefont {Wright}}]{Holl2006}%
  \BibitemOpen
  \bibfield  {author} {\bibinfo {author} {\bibfnamefont {A.}~\bibnamefont {Höll}}, \bibinfo {author} {\bibfnamefont {P.}~\bibnamefont {Maris}}, \bibinfo {author} {\bibfnamefont {C.}~\bibnamefont {Roberts}},\ and\ \bibinfo {author} {\bibfnamefont {S.}~\bibnamefont {Wright}},\ }\href {https://doi.org/https://doi.org/10.1016/j.nuclphysbps.2006.08.010} {\bibfield  {journal} {\bibinfo  {journal} {Nuclear Physics B - Proceedings Supplements}\ }\textbf {\bibinfo {volume} {161}},\ \bibinfo {pages} {87} (\bibinfo {year} {2006})},\ \bibinfo {note} {proceedings of the Cairns Topical Workshop on Light-Cone QCD and Nonperturbative Hadron Physics}\BibitemShut {NoStop}%
\bibitem [{\citenamefont {Flambaum}(2007)}]{Flambaum2007}%
  \BibitemOpen
  \bibfield  {author} {\bibinfo {author} {\bibfnamefont {V.~V.}\ \bibnamefont {Flambaum}},\ }\href {https://doi.org/10.1142/S0217751X07038293} {\bibfield  {journal} {\bibinfo  {journal} {International Journal of Modern Physics A}\ }\textbf {\bibinfo {volume} {22}},\ \bibinfo {pages} {4937} (\bibinfo {year} {2007})},\ \Eprint {https://arxiv.org/abs/https://doi.org/10.1142/S0217751X07038293} {https://doi.org/10.1142/S0217751X07038293} \BibitemShut {NoStop}%
\bibitem [{\citenamefont {Dzuba}\ and\ \citenamefont {Flambaum}(2017)}]{Dzuba2018}%
  \BibitemOpen
  \bibfield  {author} {\bibinfo {author} {\bibfnamefont {V.~A.}\ \bibnamefont {Dzuba}}\ and\ \bibinfo {author} {\bibfnamefont {V.~V.}\ \bibnamefont {Flambaum}},\ }\href {https://doi.org/10.1103/PhysRevD.95.015019} {\bibfield  {journal} {\bibinfo  {journal} {Phys. Rev. D}\ }\textbf {\bibinfo {volume} {95}},\ \bibinfo {pages} {015019} (\bibinfo {year} {2017})}\BibitemShut {NoStop}%
\bibitem [{\citenamefont {Pruttivarasin}\ \emph {et~al.}(2015)\citenamefont {Pruttivarasin}, \citenamefont {Ramm}, \citenamefont {Porsev}, \citenamefont {Tupitsyn}, \citenamefont {Safronova}, \citenamefont {Hohensee},\ and\ \citenamefont {H\"{a}ffner}}]{Haffner}%
  \BibitemOpen
  \bibfield  {author} {\bibinfo {author} {\bibfnamefont {T.}~\bibnamefont {Pruttivarasin}}, \bibinfo {author} {\bibfnamefont {M.}~\bibnamefont {Ramm}}, \bibinfo {author} {\bibfnamefont {S.~G.}\ \bibnamefont {Porsev}}, \bibinfo {author} {\bibfnamefont {I.~I.}\ \bibnamefont {Tupitsyn}}, \bibinfo {author} {\bibfnamefont {M.~S.}\ \bibnamefont {Safronova}}, \bibinfo {author} {\bibfnamefont {M.~A.}\ \bibnamefont {Hohensee}},\ and\ \bibinfo {author} {\bibfnamefont {H.}~\bibnamefont {H\"{a}ffner}},\ }\href {https://doi.org/10.1038/nature14091} {\bibfield  {journal} {\bibinfo  {journal} {Nature}\ }\textbf {\bibinfo {volume} {517}},\ \bibinfo {pages} {592–595} (\bibinfo {year} {2015})}\BibitemShut {NoStop}%
\bibitem [{\citenamefont {Derevianko}\ \emph {et~al.}(2021{\natexlab{a}})\citenamefont {Derevianko}, \citenamefont {Gibble}, \citenamefont {Hollberg}, \citenamefont {Newbury}, \citenamefont {Oates}, \citenamefont {Safronova}, \citenamefont {Sinclair},\ and\ \citenamefont {Yu}}]{Derevianko2021}%
  \BibitemOpen
  \bibfield  {author} {\bibinfo {author} {\bibfnamefont {A.}~\bibnamefont {Derevianko}}, \bibinfo {author} {\bibfnamefont {K.}~\bibnamefont {Gibble}}, \bibinfo {author} {\bibfnamefont {L.}~\bibnamefont {Hollberg}}, \bibinfo {author} {\bibfnamefont {N.~R.}\ \bibnamefont {Newbury}}, \bibinfo {author} {\bibfnamefont {C.}~\bibnamefont {Oates}}, \bibinfo {author} {\bibfnamefont {M.~S.}\ \bibnamefont {Safronova}}, \bibinfo {author} {\bibfnamefont {L.~C.}\ \bibnamefont {Sinclair}},\ and\ \bibinfo {author} {\bibfnamefont {N.}~\bibnamefont {Yu}},\ }\href {https://arxiv.org/abs/2112.10817} {\bibinfo {title} {Fundamental physics with a state-of-the-art optical clock in space}} (\bibinfo {year} {2021}{\natexlab{a}}),\ \Eprint {https://arxiv.org/abs/2112.10817} {arXiv:2112.10817 [gr-qc]} \BibitemShut {NoStop}%
\bibitem [{\citenamefont {Peik}\ \emph {et~al.}(2020)\citenamefont {Peik}, \citenamefont {Schumm}, \citenamefont {Safronova}, \citenamefont {Pálffy}, \citenamefont {Weitenberg},\ and\ \citenamefont {Thirolf}}]{Peik2020}%
  \BibitemOpen
  \bibfield  {author} {\bibinfo {author} {\bibfnamefont {E.}~\bibnamefont {Peik}}, \bibinfo {author} {\bibfnamefont {T.}~\bibnamefont {Schumm}}, \bibinfo {author} {\bibfnamefont {M.~S.}\ \bibnamefont {Safronova}}, \bibinfo {author} {\bibfnamefont {A.}~\bibnamefont {Pálffy}}, \bibinfo {author} {\bibfnamefont {J.}~\bibnamefont {Weitenberg}},\ and\ \bibinfo {author} {\bibfnamefont {P.~G.}\ \bibnamefont {Thirolf}},\ }\href {https://arxiv.org/abs/2012.09304} {\bibinfo {title} {Nuclear clocks for testing fundamental physics}} (\bibinfo {year} {2020}),\ \Eprint {https://arxiv.org/abs/2012.09304} {arXiv:2012.09304 [quant-ph]} \BibitemShut {NoStop}%
\bibitem [{\citenamefont {Derevianko}\ \emph {et~al.}(2021{\natexlab{b}})\citenamefont {Derevianko}, \citenamefont {Gibble}, \citenamefont {Hollberg}, \citenamefont {Newbury}, \citenamefont {Oates}, \citenamefont {Safronova}, \citenamefont {Sinclair},\ and\ \citenamefont {Yu}}]{FOCOS}%
  \BibitemOpen
  \bibfield  {author} {\bibinfo {author} {\bibfnamefont {A.}~\bibnamefont {Derevianko}}, \bibinfo {author} {\bibfnamefont {K.}~\bibnamefont {Gibble}}, \bibinfo {author} {\bibfnamefont {L.}~\bibnamefont {Hollberg}}, \bibinfo {author} {\bibfnamefont {N.~R.}\ \bibnamefont {Newbury}}, \bibinfo {author} {\bibfnamefont {C.}~\bibnamefont {Oates}}, \bibinfo {author} {\bibfnamefont {M.~S.}\ \bibnamefont {Safronova}}, \bibinfo {author} {\bibfnamefont {L.~C.}\ \bibnamefont {Sinclair}},\ and\ \bibinfo {author} {\bibfnamefont {N.}~\bibnamefont {Yu}},\ }\href {https://arxiv.org/abs/2112.10817} {\bibinfo {title} {Fundamental physics with a state-of-the-art optical clock in space}} (\bibinfo {year} {2021}{\natexlab{b}}),\ \Eprint {https://arxiv.org/abs/2112.10817} {arXiv:2112.10817 [gr-qc]} \BibitemShut {NoStop}%
\bibitem [{\citenamefont {Tsai}\ \emph {et~al.}(2023)\citenamefont {Tsai}, \citenamefont {Eby},\ and\ \citenamefont {Safronova}}]{SpaceQ}%
  \BibitemOpen
  \bibfield  {author} {\bibinfo {author} {\bibfnamefont {Y.~D.}\ \bibnamefont {Tsai}}, \bibinfo {author} {\bibfnamefont {J.}~\bibnamefont {Eby}},\ and\ \bibinfo {author} {\bibfnamefont {M.~S.}\ \bibnamefont {Safronova}},\ }\href {https://doi.org/https://doi.org/10.1038/s41550-022-01833-6} {\bibfield  {journal} {\bibinfo  {journal} {Nature Astronomy}\ }\textbf {\bibinfo {volume} {7}},\ \bibinfo {pages} {113} (\bibinfo {year} {2023})}\BibitemShut {NoStop}%
\bibitem [{\citenamefont {Hur}\ \emph {et~al.}(2022)\citenamefont {Hur}, \citenamefont {Aude~Craik}, \citenamefont {Counts}, \citenamefont {Knyazev}, \citenamefont {Caldwell}, \citenamefont {Leung}, \citenamefont {Pandey}, \citenamefont {Berengut}, \citenamefont {Geddes}, \citenamefont {Nazarewicz}, \citenamefont {Reinhard}, \citenamefont {Kawasaki}, \citenamefont {Jeon}, \citenamefont {Jhe},\ and\ \citenamefont {Vuleti\ifmmode~\acute{c}\else \'{c}\fi{}}}]{Hur}%
  \BibitemOpen
  \bibfield  {author} {\bibinfo {author} {\bibfnamefont {J.}~\bibnamefont {Hur}}, \bibinfo {author} {\bibfnamefont {D.~P.~L.}\ \bibnamefont {Aude~Craik}}, \bibinfo {author} {\bibfnamefont {I.}~\bibnamefont {Counts}}, \bibinfo {author} {\bibfnamefont {E.}~\bibnamefont {Knyazev}}, \bibinfo {author} {\bibfnamefont {L.}~\bibnamefont {Caldwell}}, \bibinfo {author} {\bibfnamefont {C.}~\bibnamefont {Leung}}, \bibinfo {author} {\bibfnamefont {S.}~\bibnamefont {Pandey}}, \bibinfo {author} {\bibfnamefont {J.~C.}\ \bibnamefont {Berengut}}, \bibinfo {author} {\bibfnamefont {A.}~\bibnamefont {Geddes}}, \bibinfo {author} {\bibfnamefont {W.}~\bibnamefont {Nazarewicz}}, \bibinfo {author} {\bibfnamefont {P.-G.}\ \bibnamefont {Reinhard}}, \bibinfo {author} {\bibfnamefont {A.}~\bibnamefont {Kawasaki}}, \bibinfo {author} {\bibfnamefont {H.}~\bibnamefont {Jeon}}, \bibinfo {author} {\bibfnamefont {W.}~\bibnamefont {Jhe}},\ and\ \bibinfo {author} {\bibfnamefont {V.}~\bibnamefont {Vuleti\ifmmode~\acute{c}\else \'{c}\fi{}}},\ }\href
  {https://doi.org/10.1103/PhysRevLett.128.163201} {\bibfield  {journal} {\bibinfo  {journal} {Phys. Rev. Lett.}\ }\textbf {\bibinfo {volume} {128}},\ \bibinfo {pages} {163201} (\bibinfo {year} {2022})}\BibitemShut {NoStop}%
\end{thebibliography}%

\end{document}